\documentclass{article}
\usepackage{epsf}
\begin{document}
\title{Prediction of the In-Situ Dust Measurements of the Stardust
Mission to Comet 81P/Wild 2}
\author{M. Landgraf\\ NASA/Johnson Space Center, Houston, TX, USA,\\
        M. M\"uller, and E. Gr\"un\\ Max--Planck--Institut f\"ur
        Kernphysik, Heidelberg, Germany} 
\date{}
\maketitle
\section*{Abstract}
We predict the amount of cometary, interplanetary, and interstellar
cosmic dust that is to be measured by the Com\-et\-ary and Interstellar
Dust Analyzer (CIDA) and the aerogel collector on-board the
Stardust spacecraft during its fly-by of comet P/Wild 2 and during the
interplanetary cruise phase. We give the dust flux on the spacecraft
during the encounter with the comet using both, a radially symmetric
and an axially symmetric coma model. At closest approach, we predict a
total dust flux of $10^{6.0}\;{\rm m}^{-2}\;{\rm s}^{-1}$ for the
radially symmetric case and $10^{6.5}\;{\rm m}^{-2}\;{\rm s}^{-1}$ for
the axially symmetric case. This prediction is based on an observation of
the comet at a heliocentric distance of $1.7\;{\rm AU}$. We reproduce
the measurements of the Giotto and VEGA missions to comet P/Halley
using the same model as for the Stardust predictions.
The planned measurements of {\em interstellar} dust by Stardust 
have been triggered by the discovery of interstellar
dust impacts in the data collected by the Ulysses and Galileo dust
detector. Using the Ulysses and Galileo measurements we predict that
$25$ interstellar particles, mainly with masses of about $10^{-12}\
{\rm g}$, will hit the target of the CIDA experiment. The
interstellar side of the aerogel collector will contain $120$
interstellar particles, $40$ of which with sizes greater than $1\ {\rm
\mu m}$. We furthermore investigate the ``contamination'' of the CIDA
and collector measurements by interplanetary particles during the
cruise phase.
\section{Introduction}
\begin{figure}[htb]
\hspace{-5mm}
\epsfbox{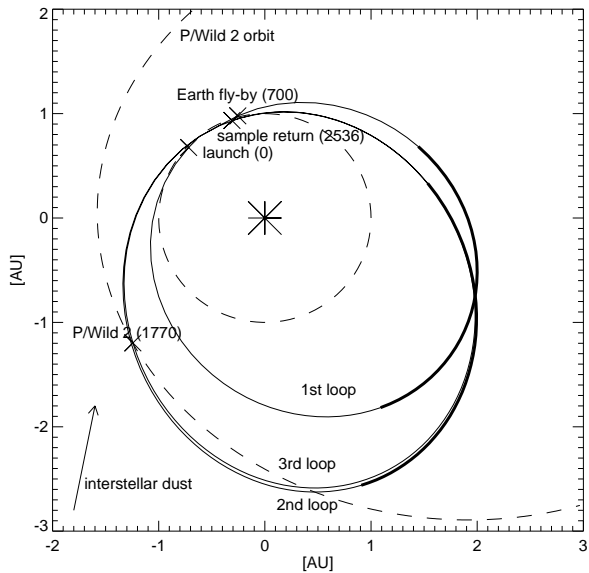}
\caption{\label{fig_traj} The trajectory of Stardust (solid line) in
the ecliptic plane (the verenal equinox direction is to the
right). The orbits of P/Wild 2 and Earth are indicated as dashed
lines. Mission events are marked with the time given in days after
launch in brackets. The flow direction of interstellar particles from
the LIC as derived from the Ulysses/ Galileo measurements is shown by
the arrow in the lower left corner. Thick lines indicate the part of
the orbit used to collect interstellar dust with the Aerogel
collector.}
\end{figure}

The Stardust mission to comet 81P/Wild 2 is dedicated to the in situ
measurement and sample return of cosmic dust. At the comet the
pristine cometary material is investigated which is assumed to be a
good sample of the nebula from which the Solar System has
formed. Pre-solar interstellar grains that survived the formation
process may therefore be present in the samples collected or measured
in situ. On the way to the com\-et the opportunity is taken to
investigate contemporary interstellar dust that traverses the Solar
System and originates from the local interstellar low-density diffuse
cloud which crosses the Sun's way through the Milky Way
\cite{frisch95}. We predict the dust flux on the two main instruments
on-board Stardust, the aerogel dust collector and the Cosmic and
Interstellar Dust Analyzer (CIDA) \cite{brownlee97}. The dust
collector is a plate with an area of $\approx 0.1\;{\rm m}^2$ that is
to be exposed to the dust stream. Particles hitting the collector are
trapped inside the aerogel \cite{albee94}. CIDA is an in-situ impact
plasma dust detector with a time-of-flight mass spectrometer similar
to the PUMA/PIA instruments flown on Giotto \cite{kissel86}.
Stardust launches in February 1999 and flies by Earth in January
2001 for gravity assist. The slow ($6.1\ {\rm km}\ {\rm s}^{-1}$)
fly-by of comet P/Wild 2 occurs on January 1st, 2004 at a heliocentric
distance of $1.86\ {\rm AU}$ when the comet is on the out-bound part of
its orbit. During the fly-by the distance of the
spacecraft to the comet nucleus is $\approx 100\ {\rm km}$, and the
phase angle, i.e. the Sun--comet--spacecraft angle, is
$70^\circ$. After fly-by Stardust returns to Earth in January 2006,
where the sample-return capsule that contains the aerogel collector is
separated from the spacecraft and sent to a direct entry into Earth's
atmosphere. The collection and in situ measurement of interstellar
dust occurs during the cruise phase as shown in figure
\ref{fig_traj}.
During the short time of the encounter with the comet cometary dust
particles are collected which are lifted from the comet's nucleus and
accelerated into the coma by the drag of the outflowing gas. The
amount of material in a comet's coma depends on the activity of its
nucleus, which, in turn, depends on the distance from the Sun and on
the properties of the nucleus' surface. To determine P/Wild 2's
activity we use ground based observations. For the properties of the
dust phase, especially the mass distribution of dust particles,
contained in a comet's coma, we rely on data collected with the VEGA
and Giotto spacecraft at comet P/Halley during its visit of the inner
Solar System in 1986. Since P/Halley and P/Wild 2 differ in activity
and size, and since the VEGA and Giotto measurements had a different
geometry than the measurements of Stardust, we have to construct 
a  model  of  the  coma  and  use 
the  measured  activity, brightness and mass
distribution as an input to this model. By using the model to
reproduce the spacecraft measurements of the dust fluence at P/Halley,
we can validate the model and the procedure of taking the input
activity and brightness from ground based observations.
The secondary goal of the Stardust mission is the collection and
in-situ measurement of contemporary interstellar dust particles which
enter the Solar System from the local interstellar cloud (LIC). The
same instruments are used for the interstellar dust measurements as
for the cometary part of the mission. During definite parts of
Stardust's cruise, the dust collector will expose its backside to the
interstellar dust stream and on other occasions CIDA will be pointed
into the upstream-direction (see figure \ref{fig_traj}).  The
interstellar dust stream was discovered by
Gr\"un et al. (1993) \cite{gruen93} in the data collected by the
dust-detector on-board the 
Ulysses spacecraft after fly-by of Jupiter. It was clearly identified
and distinguished from interplanetary dust by its opposite impact
direction, its impact velocity in exceess of the escape velocity at
Jupiter distance, and its constant impact rate at high heliocentric
latitudes. The Ulysses measurements have been confirmed by impacts
from the retrograde direction measured by the  identical
dust-detector 
 on-board  the  Galileo  spacecraft
\cite{baguhl95b}. The total flux of
interstellar particles was determined to be $1.5\cdot 10^{-4}\ {\rm
m}^{-2}\ {\rm s}^{-1}$
\cite{gruen94}. The mass distributions of the Ulysses and Galileo
measurements have been found to be nearly identical and range from
$10^{-15}\ {\rm g}$ to $10^{-8}\ {\rm g}$ with particles of masses of
$10^{-13}\ {\rm g}$ being most abundant
\cite{landgraf98a}. From the
impact direction it was derived that interstellar dust particles come
from the same direction \cite{baguhl95a} as interstellar neutral
Helium which was measured by the Ulysses/GAS experiment
\cite{witte93}. The upstream-direction of the interstellar Helium was
determined to be $\lambda_{\infty,{\rm gas}}=254.7^\circ \pm
1.3^\circ$ (heliocentric longitude) and $\beta_{\infty,{\rm
gas}}=4.6^\circ\pm 0.7$ (heliocentric latitude)
\cite{witte96}. Furthermore the relative velocity of the gas was
given by $v_{\infty,{\rm gas}}=25.4\pm 0.5\ {\rm km}\ {\rm
s}^{-1}$. These parameters relate the interstellar material measured
in the Solar System with the LIC which has been identified in the
Doppler shift measured in the UV by HST-GHRS
\cite{lallement92}. The Doppler-shift indicates a relative motion between
the Sun and the LIC with a velocity of $25.7\pm 0.5\;{\rm km}\;{\rm
s}^{-1}$.
\begin{figure}[htb]
\epsfxsize=.9\hsize
\centering\epsfbox{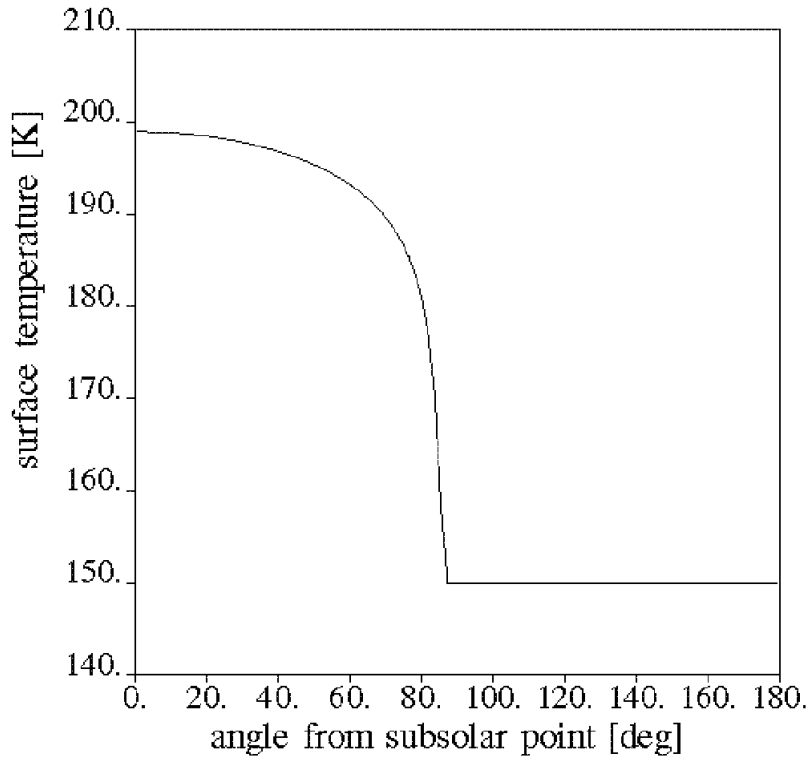}
\caption{\label{srtmp}
Equilibrium surface temperature versus the angle $\vartheta_s$ from
the sub-solar point as given by the solution of equation
(\ref{surt}).}
\end{figure}

In this paper we use the expressions ``flux'' for the number of impacting
particles of a given mass per unit area and time, ``fluence'' for the flux
integrated over time, ``total flux'' for the flux of particles of all
masses, and ``total fluence'' for the total flux integrated over time.

\section{Coma Model\label{commod}}

In the following we describe the coma model we use to determine the
cometary dust flux on the Stardust spacecraft. Because there is no
standard model of the gas and dust environment of a comet nucleus,
yet, we describe the assumptions and calculations we use in
detail. So the reader can see how different assumptions or values of
parameters change the results. 

Since the dust is dragged into the coma by outflowing gas, we first
have to calculate the gas density and velocity inside the coma. We
assume that the outflow is symmetric about the Sun-nucleus
axis. Therefore, all spatial distributions depend only on the angle
$\vartheta_s$ between the considered location and the axis of symmetry
and the radial distance $r$. We furthermore assume that
the dust particles do not affect the gas flow, i.e. we use a
test-particle approach.

\subsection{Gas Phase}

To calculate the state of the gas around the nucleus, we need to
 know  the  activity  distribution  on
 the  nucleus'  surface, which is governed
by the temperature distribution.

The temperature at a location on the day side of the nucleus' surface
is calculated such that there is an equilibrium between solar
illumination, water sublimation, and blackbody reradiation. Therefore,
the surface temperature $T_s(\vartheta_s)$ is the solution of the
following equation (the meaning and values of the
parameters in the next two equations are given in table
\ref{surtab}). 
\begin{equation}\label{surt}
\renewcommand{\arraystretch}{1.5}
\begin{array}{l}
\left( 1 - A_B \right) \left( \frac{f_{Sun}}{(r_h/1AU)} \right)^2
\cdot \cos( \vartheta_s ) \\
\hspace{2cm} = \frac{L}{N_A} Z_{\rm HK}(T_s(\vartheta_s))
+ \epsilon \left( \sigma(T_s(\vartheta_s)) \right)^4,
\end{array}
\renewcommand{\arraystretch}{1}
\end{equation}
where $Z_{\rm HK}(T_s)$ is the Hertz--Knudsen sublimation
rate given by:
\begin{eqnarray}\label{eqn_zhk}
Z_{\rm HK}(T_s) & = & \frac{ p(T_s) }{ \sqrt{2\pi m_{\rm H_2O} k T_s} }
\end{eqnarray}
We use the approximation $p(T_s)=Ae^{(-B/T_s)}$ for the vapor pressure
\cite{fanale}.
\begin{table}
\caption{\label{surtab}
Parameters for computation of the surface
temperature.}
\begin{tabular}{ll}
\hline
Solar constant &$f_{\rm Sun}=1370\;{\rm W}\;{\rm m}^{-2}$\\
nucleus surface albedo\footnotemark[1]  & $A_B=6\%$\\
Emissivity of comet surface &$\epsilon=1$\\
Boltzmann constant & $k=1.38\cdot 10^{-23}\;{\rm J}\;{\rm K}$\\
Stefan Boltzmann constant & $\sigma=5.67\cdot 10^{-7}\;{\rm W}\;{\rm
K}^{-4}\;{\rm m}^{-2}$\\
Advogadro Number   & $N_A=6\cdot 10^{23}{\rm mol}^{-1}$ \\
Latent heat of water ice & $L=36\;{\rm kJ}\;{\rm mol}^{-1}$\\
Vapor pressure parameter & 
$A=3.56\cdot 10^{12}\;{\rm N}\;{\rm m}^{-2}$\\
Vapor pressure parameter & $B=6141\;{\rm K}$ \\
\hline
{\footnotesize $^1$ bolometric Bond albedo}
\end{tabular}
\end{table}
On the night side of the nucleus the temperature is
constantly set to $150\;{\rm K}$. This value is also used at the day
side near the terminator if the solution of equation (\ref{surt}) is
smaller than $150\;{\rm K}$. Figure \ref{srtmp} shows the equilibrium
temperature distribution determined from equation (\ref{surt}).
\begin{figure}[htb]
\epsfxsize=.9\hsize\epsfbox{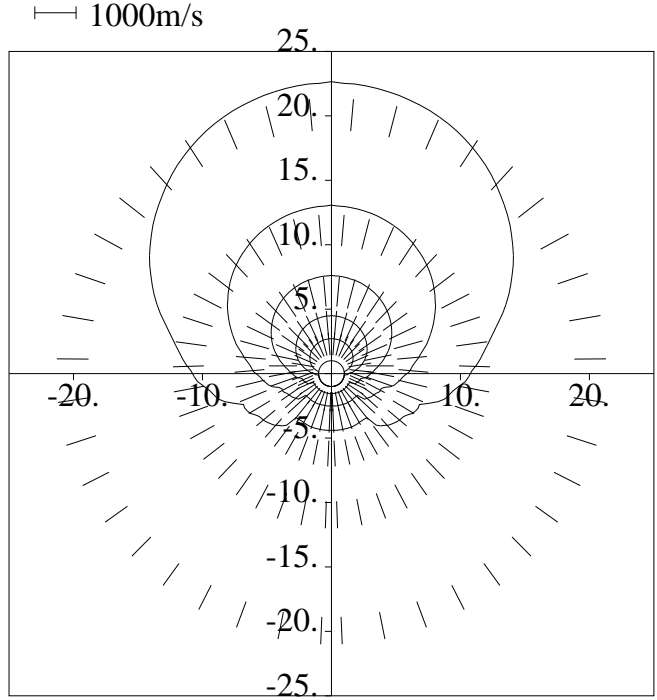}
\caption{\label{gasiso} Solid lines represent lines of constant gas
density (arbitrary units) in the region up to 25 comet radii
($=50\;{\rm km}$) around the nucleus. The Sun direction is upward and
the short radial lines indicate the gas velocity (the scale on the top
can be used to determine the modulus of the velocity). The gas density
above the comet night side (bottom) is much lower than the density
above the day side. However, the gas velocity above the day and night
side is comparable. Apart from a region above the terminator, the gas
flow is almost radial close to the nucleus.}
\end{figure}

Now, we obtain the activity $Q_{\rm HK}$ of the comet by inserting the
temperature distribution into equation (\ref{eqn_zhk}) and integrate
over $\vartheta_s$.
\begin{eqnarray}
Q_{\rm HK} & = & 2\pi r_n^2 \int_{-1}^{1} Z_{\rm HK}(\vartheta_s)
d\cos(\vartheta_s).
\end{eqnarray}
In our model we use the outgasing activity of the nucleus given by
Farnham and Schleicher (1997) \cite{obs1}. On 1997 March 5 they report an
$\rm OH$-activity $Q_{\rm OH}=5.9\cdot 10^{27}\;{\rm s}^{-1}$ and a value of
the albedo--filling factor--radius product of $Af\rho=4.27\;{\rm
m}$. At this time the
comet was at a heliocentric distance of $1.7\;{\rm AU}$ and was
observed at a phase angle of $27^\circ$. Although the observation was
taken on the in-bound part of the comet's orbit, and the Stardust
spacecraft will approach the comet on the out-bound part, we assume
that the observed activity is comparable to the comet's activity
during the encounter. Therefore, we run our coma model for a comet at
$1.7\;{\rm AU}$ with a water activity equal to the observed
$\rm OH$-activity $Q_{\rm H_2O}=5.9\cdot 10^{27}\;{\rm s}^{-1}$. To match the
observed activity, the water sublimation rate is
calculated by scaling the Hertz--Knudsen sublimation rate:
\begin{eqnarray}
Z_{\rm H_2O}(\vartheta_s) & = & \frac{Q_{\rm H_2O}}{Q_{\rm HK}} \cdot Z_{\rm
HK}(T_s(\vartheta_s))
\end{eqnarray}
In addition to the water activity, we assume a $\rm CO$-activity of $10\%$
of the water activity, i.e. $Q_{\rm CO}=5.9\cdot 10^{26}\;{\rm s}^{-1}$,
and a constant distribution above the nucleus' surface
$Z_{\rm CO}(\vartheta_s)=Q_{\rm CO}/(4\pi r_n^2)$. For calculating the
acceleration of the dust particles by the cometary gas, we have to
consider the most abundant  gas  molecules  only,
 therefore  other  constituents  than
$\rm H_2O$ and $\rm CO$ are not considered in the model and
the total gas activity is set to $Q_{\rm gas}=Q_{\rm H_2O}+Q_{\rm CO}=6.5\cdot
10^{27}\;{\rm s}^{-1}$. Accordingly, we set the total gas activity
distribution to $Z_s(\vartheta_s) = Z_{\rm H_2O}(\vartheta_s) +
Z_{\rm CO}(\vartheta_s)$. Using $Z_s(\vartheta_s)$, the thermodynamic
state of the gas can be determined by a numerical procedure described
by Crifo et al. (1995) \cite{crifo2} and Knollenberg (1993) \cite{knoll}.

As an input to the calculation of the gas state we use a mean mass of
a gas molecule of $m_{\rm gas} = (m_{\rm H_2O} Q_{\rm H_2O} + m_{\rm
CO} Q_{\rm CO})$ $/
(Q_{\rm H_2O} + Q_{\rm CO})$ and a mean degree of freedom per molecule of
$f_{\rm gas} = (f_{\rm H_2O} Q_{\rm H_2O} + f_{\rm CO} Q_{\rm CO}) /
(Q_{\rm H_2O} +
Q_{\rm CO})$. This simplification introduces some physical inconsistency
into the model, but the results do not depend significantly on these
choices.

Another important parameter for modeling the gas outflow is the initial gas
temperature above the nucleus' surface, because it determines the 
velocity reached by the gas, and therefore also has an impact on
the dust acceleration. The initial gas temperature above the nucleus'
surface does not necessarily coincide with the surface temperature, because
the gas is not at rest with respect to the surface. This
effect was taken into account as it is described in  
Crifo and Rodionov (1997) \cite{crifo1} (Appendix B). 

As shown in figure \ref{gasiso}, the gas density determined by
our thermodynamic  model  is  distributed 
highly  asymmetric around the
nucleus. On the day side much more gas is available to accelerate dust
from the surface. The flow direction is nearly radial. In the outer
region of the coma, the gas is not dense enough to exert a significant
drag force on the dust particles. We assume that the gas drag is
negligible outside a maximum distance $r_{\rm max}$ of $25$ times the
radius of the nucleus. After leaving this region, the dust particles
move independently from the gas.

\subsection{Dust Phase}
The dust phase of a comet's coma consists of all dust particles lifted
and accelerated from the nucleus by the gas drag.  The mass range of
dust particles emitted by a comet covers many orders of magnitude and
the dynamics of these particles depend on their mass. Therefore we
divide the mass range $[10^{-20}\;{\rm kg},10^{0}\;{\rm kg}]$ into
$20$ logarithmic mass decades $[m_{i-1},$ $m_i]$, $i=1,\ldots,20$ which
we call ``dust classes'' in the following. All dust particles which
are released with a mass inside one of the intervals $i$ are modeled
by a dust particle of a representative mass $m_{d,i}$. We assume that
no processes which change the properties of a dust particle,
e.g. fractionation due to mutual collisions, take place after a
particle has left the nucleus' surface. Furthermore, we assume the
dust particles to be spherical with a constant bulk density
$\rho_d=1\;{\rm g}\;{\rm cm}^{-3}$ and a radius
$s_i=(3m_{d,i}/(4\pi\rho_d))^{(1/3)}$. For the determination of the
amount of dust in the coma, we derive a dust-to-gas mass ratio
$\chi_i$ for each dust class. Using the dust to gas ratios $\chi_i$,
we calculate the number of dust particles released per unit area and
time.
\begin{eqnarray}
Z_{d,i}(\vartheta_s) & = & \chi_i \frac{m_{\rm gas}}{m_{d,i}}Z_{\rm
gas}(\vartheta_s)
\end{eqnarray}
We take into account that particles cannot be lifted from the comet
surface above a given mass by setting the dust activity to zero at
locations on the comet surface where the gravitational attraction of
the comet nucleus exceeds the drag force exerted by the gas.

To calculate the gravitational force of the nucleus, we determined the
nucleus' size by using ground-based observations of P/Wild 2 at large
heliocentric distances. Fitzsimmons and Cartright (1995) \cite{obs2}
report a red magnitude of $R=22$ at the heliocentric distance of
$r_h=4.4\;{\rm AU}$, a geocentric distance of $\Delta=4.0\;{\rm AU}$,
and a phase angle of $\alpha=12^\circ$. We assume that P/Wild 2's
nucleus has a spherical shape. The radius of the nucleus $r_n$ can
be calculated by (compare Jewitt (1991) \cite{jew})
\begin{eqnarray}\label{eqrn}
r_n & = & r_h \frac{\Delta} {1\;{\rm AU}} \sqrt{ \frac{10^{-0.4 (R -
R_{\rm Sun})}} {p_{\rm nuc} \cdot j_{\rm nuc}(\alpha)} },
\end{eqnarray}
where $R_{\rm Sun}=-27.22$ is the red magnitude of the Sun \cite{land},
$p_{\rm nuc}=4\%$ is used for the geometric albedo of the nucleus and
$j_{\rm nuc}(\alpha)=10^{-0.4\delta\alpha}$ is the phase function of the nucleus 
with $\delta=0.035\;{\rm deg}^{-1}$ \cite{jew}.
We find $r_n=2.3\;{\rm km}$. Because
Fitzsimmons and Cartright (1995) \cite{obs2}
reported a ''near--stellar appearance'', there was already some
contribution of the dust coma to the observed intensity. Therefore,
$r_n=2.3\;{\rm km}$ is an overestimation of the radius of the
nucleus and we use the value $r_n=2\;{\rm km}$ in our model.
\begin{figure}[htb]
\epsfxsize=.8\hsize\epsfbox{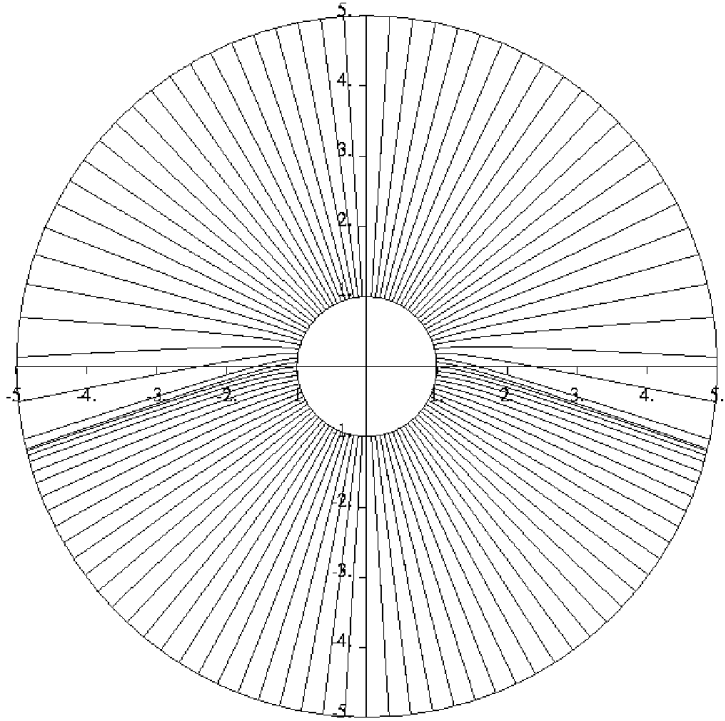}
\caption{\label{dsttra09}
Trajectories of dust particles with radii of $8.8\;{\rm \mu m}$, which
start at equidistant positions on the surface. The scales in the image
are nucleus radii. The velocity of the particles is
dominated by the radial component, far from the nucleus the tangential
velocity component is negligible. The distance of the trajectories
above the terminator is increased due to the tangential gas flow in
this region. }
\end{figure}

\begin{figure}[htb]
\epsfxsize=.8\hsize\epsfbox{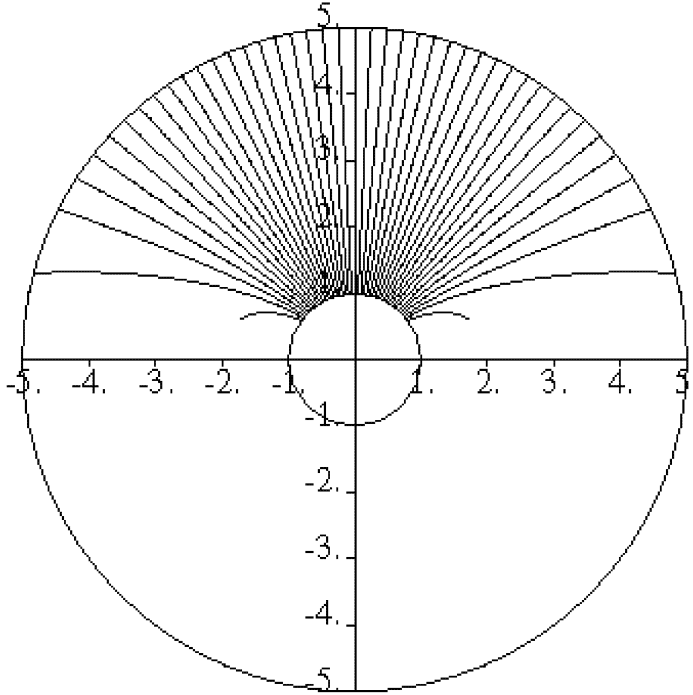}
\caption{\label{dsttra18}
Trajectories of dust particles with a radius of $8.6\;{\rm
mm}$. These big
dust particles can only be lifted in a region around the sub-solar
point. Near the boundary of this region there are trajectories which
start to fall back in nucleus direction. In our model we neglect the
contribution of these dust particles to the dust number density of the
coma.}
\end{figure}

Assuming a bulk density of $\rho_n=0.5\;{\rm g}\;{\rm cm}^{-3}$, the
gravitational acceleration is given by:
\begin{eqnarray}
\vec{a}_{\rm grav} & = & \mu_n
\frac{\vec{r}} {\left|\vec{r} \right|^3}
\hspace{0.5cm} \mbox{with} 
\hspace{0.5cm} \mu_n=1118\;{\rm m}^3\;{\rm s}^{-2}.
\end{eqnarray}
The acceleration of a dust particle due to gas drag
can be expressed using the drag coefficient $C_D$:
\begin{eqnarray}
\vec{a}_{\rm drag} & = & \frac{1}{2} m_{\rm gas} n_{\rm gas} \cdot
\frac{\pi s_i^2}{m_{d,i}} C_D \left| \vec{v}_{\rm gas} 
- \vec{v}_{\rm dust} \right| \nonumber \\
&& \cdot \left( \vec{v}_{\rm gas} - \vec{v}_{\rm dust} \right)
\end{eqnarray}
The drag coefficient in the free molecular approximation is given by
Probstein (1968) \cite{prob}
\begin{eqnarray} 
C_D & = & \frac{2 \sqrt{\pi}} {3s} \sqrt{\frac{T_d}{T_{\rm gas}}} +
\frac{2s^2+1} {s^3 \sqrt{\pi}} e^{(-s^2)} \nonumber \\
&& + \frac{4s^4+4s^2-1} {2s^4}
\mbox{erf}(s),
\end{eqnarray}
where $s=|\vec{v}_{\rm gas} - \vec{v}_{dust}| /\sqrt{2kT_{\rm gas} /
m_{\rm gas}}$.  The dust temperature is set to $T_d=228\;{\rm K}$
(compare Divine (1981) \cite{div1}: $T_d=310\;{\rm K} \cdot (r_h/1\;{\rm
AU})^{-0.58}$ and $r_h = 1.7\;{\rm AU}$).  At every point on the
nucleus' surface for which the gas drag acceleration exceeds the
gravitational acceleration of the nucleus, the equation of motion of a
dust particle can be solved numerically inside the gas flow (see
figure \ref{dsttra09} and
\ref{dsttra18}). For large particles which can be lifted only from
the comet's day side, there are trajectories for which the radial
velocity is not directed away from the nucleus but fall back into the
nucleus' direction (see figure \ref{dsttra18}).  Since this behavior
is only found in a narrow region, we do not consider these particles
in the following.

Apart from a narrow region near back-falling particles, the dust
velocity is dominated by the radial velocity component. Since the
contribution of the tangential velocity to the total dust velocity
decreases with increasing distance from the nucleus' surface, the dust
trajectories can be approximated by straight radial lines with
constant radial velocity at large distances. Outside the maximum
distance $r_{\rm max}$ we approximate the dust trajectories by
straight, radial lines and constant velocities $v_{e,i}(\vartheta)$
which is equal to the radial velocity computed for a dust particle at
the distance $r_{\rm max}$. The angle $\vartheta$ denotes the angle of
the position of the dust particle with respect to the Sun direction at
$r_{\rm max}$, and is also the angle of the extrapolated trajectory
with the Sun direction.

The dust number density outside the radius $r_{\rm max}$ can then be
computed using the position $\vartheta_s$ of the trajectory at the
nucleus' surface as a function of the angle $\vartheta$ with respect
to the Sun direction. To calculate the number density $n_{d,i}$ at
positions with $r>r_{\rm max}$ and angle $\vartheta$, we introduce
the activity distribution of the comet outside the acceleration zone:
\begin{eqnarray}
Z_{e,i} & = & r_n^2 Z_s(\vartheta_s(\vartheta)) \frac{d
\cos(\vartheta_s)} {d \cos(\vartheta)}
\end{eqnarray}\label{numb}
For the number density $n_{d,i}$ we get:
\begin{eqnarray}
n_{d,i}(r,\vartheta) & = & \frac{Z_{e,i}(\vartheta)}{r^2 v_{e,i}(\vartheta)}
\end{eqnarray}
The number density inside the acceleration cone can be calculated in
an analogous way. 

\begin{figure}[htb]
\hspace{3mm}\epsfxsize=\hsize\epsfbox{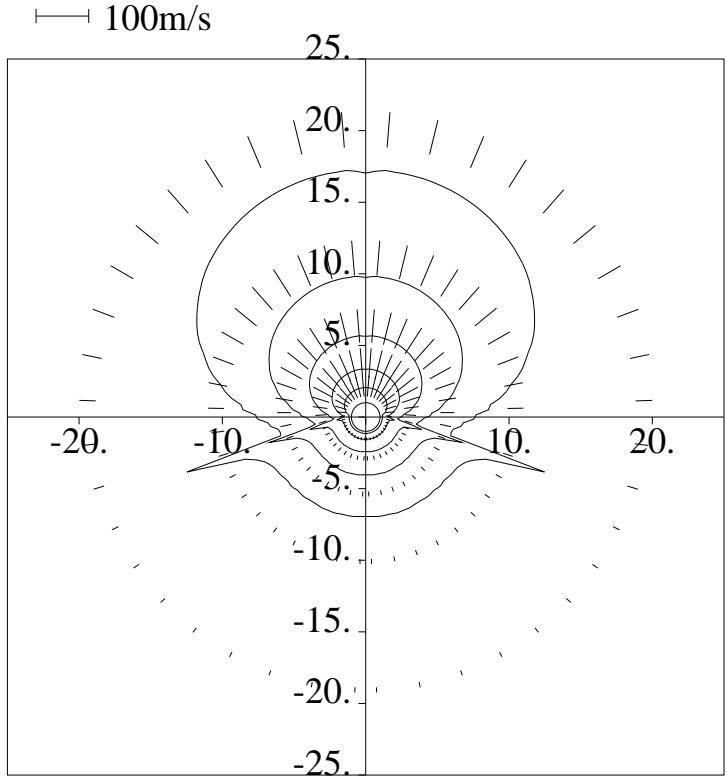}
\caption{\label{dstiso}
The lines of constant dust number density of particles with radii of
$8.8\mu m$ in the region within 25 radii of the nucleus. The
radial lines indicate the velocity of the dust particles. Because of
the low gas density, the dust particle velocity is smaller above the
night side.  Due to the increased tangential
distance of the dust trajectories, the dust number
density is small above the terminator (see figure \ref{dsttra09}).}
\end{figure}

As a result, we show the contours of constant dust particle density in
figure \ref{dstiso}. The region of high dust density occupies a larger
volume on the day side than on the night side. Due to the drop in
activity at the terminator, the dust particle density right above the
terminator is lower than on the night side. Dust particles from the
day side can propagate freely to the night side which extends the
region of high dust particle density just behind the terminator in a
narrow radial feature. If this is a real feature at a comet and if it
can be observed in reality remains an open question. The dust particle
density opposite to the Sun direction is the result of the night side
activity.

We want to use the model to determine the dust flux on the Stardust
instruments during the fly-by of the nucleus. We can neglect
the influence of radiation pressure on the particle's trajectories,
 because  Stardust  approaches  the 
nucleus sufficiently close. Because
the velocity of Stardust relative to nucleus is much larger than the
velocity of the dust particles, the dust impact velocity at the
spacecraft is dominated by the spacecraft's motion. We therefore
calculate the dust flux on the spacecraft by the product of the dust
number density and the spacecraft velocity $v_{\rm SC}$ relative to
the nucleus.
\begin{eqnarray}\label{fsc}
f_{\rm SC} & = & n_{d,i} \cdot v_{\rm SC}
\end{eqnarray}
From the result in equation (\ref{fsc}) we can derive the dust fluence
on the instruments, which is given by the time integral over the flux
and represents the column density of dust particles along the
spacecraft's trajectory. We need the column density of dust particles
also to calculate the brightness measured by an ground-based observer
for a given line of sight.

As we show in Appendix A, the column  density (in number of
dust particles per ${\rm m}^2$) for an image plane with phase angle
$\alpha$ can be calculated by
\begin{eqnarray}\label{nproj}
\frac{d n_{d,i}}{d(x',y')} & = & \frac{1} {\rho} \int_{-u}^{u}
d\cos(\vartheta) \; \frac{Z_{e,i}(\vartheta)} {v_{e,i}(\vartheta)}
\nonumber \\
&&\cdot \frac{1} {\sqrt{u^2 - \left(\cos(\vartheta)^2\right)}}
N(\alpha,\eta,\vartheta),
\end{eqnarray}
where $u = u(\alpha,\eta) = \sqrt{1 - (\sin(\eta) \cdot
\sin(\alpha))^2}$. The variables $\rho$ and $\eta$ represent the
position of the line of sight in the projected plane: $\rho$ is the
distance from the (projected) nucleus position and $\eta$ denotes the
direction of the line of sight position with anti-solar direction
projected in the image plane (compare figure \ref{app1}).
$N(\alpha,\eta,\vartheta)$ is equal to the number of intersections of
a line of sight with the cone of constant $\vartheta$ and is equal to
$0$, $1$, or $2$ (see table \ref{naet}).
%
%
\begin{table}
\caption{\label{naet} Number of intersections of a line of sight of
angle $\eta$ with a cone of opening angle $\vartheta$ for different cases as
explained in the text. Note that $(N(\alpha, \eta, \vartheta) +
N(\alpha, \pi-\eta, \vartheta))/2 = 1$ for $|\cos(\vartheta)| < u$.}
\begin{tabular}{lll}
\hline
Condition 1 & Condition 2 & $N(\alpha,\eta,\vartheta)$ \\
\hline
\hline
$u < \cos(\vartheta)$ & none & $0$ \\
\hline
$|\cos(\alpha)|<\cos(\vartheta)<u$ & 
\begin{tabular}{l}
$\cos(\eta)<0$\\
$\cos(\eta)>0$
\end{tabular}
&
\begin{tabular}{@{\hspace{0cm}}l@{\hspace{0cm}}}
$2$\\
$0$
\end{tabular}
\\
\hline
$\begin{array}{l}
|\cos(\alpha)|>|\cos(\vartheta)>\\
-|\cos(\alpha)| \end{array} $ & none & 1 \\
\hline
$-|\cos(\alpha)|>\cos(\vartheta)>-u$ & 
\begin{tabular}{l}
$\cos(\eta)<0$\\
$\cos(\eta)>0$
\end{tabular}
&
\begin{tabular}{@{\hspace{0cm}}l@{\hspace{0cm}}}
$0$\\
$2$
\end{tabular}
\\
\hline
$-u>\cos(\vartheta)$ & none & $0$ \\
\hline
\end{tabular}
\end{table}

For comparison with our axially symmetric model we provide the column
density in the radially symmetric case:
\begin{eqnarray} \label{nprojrsym}
\frac{d n_{d,i}}{d(x',y')} & = & \frac{\pi Z_{e,i}}{\rho v_{e,i}}.
\end{eqnarray}

\subsection{Adjustment of the Dust Activity to the Observed $Af\rho$-value}
We adjust the dust activity of our model so that the observed
$Af\rho_{\rm obs}=4.27\;{\rm m}$ is matched. The $Af\rho$-parameter is related to
the intensity $f_{\rm dust}$ received from dust particles inside a field
of view with radius $\rho$ by A'Hearn et al. (1984) \cite{ahearn}:
\begin{eqnarray}
Af\rho & = & 4\frac{\Delta^2(r_h/1AU)^2} {f_{\rm Sun}} \cdot
\frac{f^*_{\rm dust}} {\rho},
\end{eqnarray}
where $f^*_{\rm Sun}$ is the radiation flux density of the Sun at a
heliocentric distance of $1\;{\rm AU}$ observed in the same band of
the spectrum as the comet, $\Delta$ is the geocentric distance and
$r_h$ the heliocentric distance.

The intensity received due to a single dust particle is given by
\begin{eqnarray}
I_{sng,i} & = & pj(\alpha) \cdot \frac{\pi s_i^2}{\pi \Delta^2} \cdot
\frac{f^*_{\rm Sun}} {(r_h/1AU)^2},
\end{eqnarray}
where $p$ is the dust geometric albedo and $j(\alpha)$ is the
dust phase function of the phase angle $\alpha$.
Using the number of dust particles inside the aperture $N_{d,i}$ 
(see equation (\ref{naper})), we can determine
the $Af\rho$-parameter of our model by
\begin{eqnarray}
Af\rho_{\rm mod} & = & 4 pj(\alpha) \sum_i s_i^2 \int_{0}^{2\pi}
\int_{-u}^{u} \frac{Z_{e,i}(\vartheta)} {V_{e,i}(\vartheta)}\nonumber
\\
&& \cdot
\frac{1} {\sqrt{u^2 - \cos^2(\vartheta)}} d\cos(\vartheta)\; d\eta.
\end{eqnarray}
Here we assume that all dust particles have a geometric albedo 
and a phase function independent of size. For the geometric albedo we assume
a value of $4$\% and take the phase function from Divine (1981) \cite{div1}.

To match the observed $Af\rho_{\rm obs}$ with the coma model, we have to
scale the dust activity of the comet, i.e. the dust to gas ratio of
all dust classes have to be scaled by the factor
$Af\rho_{\rm obs}/Af\rho_{\rm mod}$.

\subsection{Dust Mass Distribution}
The mass distribution of the dust particles is the most crucial
parameter of the model, because a change in the mass distribution can
change the total dust fluence by orders of magnitude. Therefore we take
special care to choose a reasonable mass distribution. In this
section we explain our choice and discuss the resulting fluence in
comparison with the measured fluence in the coma of P/Halley. The
reader might want to check what the effects of a different choice of
the mass distribution are.

The VEGA spacecraft measured the number of particles per unit area as
a function of mass. We use this measurement as the mass distribution
of our model.  As an analytical representation of the mass distribution,
we use a fit by Divine and Newburn (1987) \cite{div2}.
\begin{eqnarray}\label{msd}
F(m) & = & F_t \left(\frac{(1+x)^{b-1}} {x^b}\right)^{ac}
\hspace{0.5cm} \\
&& \mbox{with} \hspace{0.5cm}
x = \left(\frac{m} {m_t}\right)^{1/c}. \nonumber
\end{eqnarray}
The parameters which fit the cumulative mass distribution of the VEGA
2 fluence best are $a=0.9$, $b=0.29$, $c=2.16$, and $m_t=1.6\cdot
10^{-13}\;{\rm kg}$ \cite{div2}. Since the mass
distribution is only used to determine the relative but not the total
abundance of particles in different dust classes, $F_t$ can be set to
an arbitrary value.  Note that $-a=-0.9$ is the exponent of the
cumulative mass distribution for big dust particles ($m \gg m_t, x
\gg 1$), and $-ac = -1.9$ is the exponent for small particles ($m \ll
m_t, x \ll 1$).

For each dust class $[m_{i-1},m_i],\; i=1,\ldots,20$ considered in the
model, the fluence on the VEGA 2 spacecraft inside one of the mass
bins is proportional to
\begin{eqnarray}
n_{VG2,i} & \propto & F(m_{i-1})-F(m_i)
\end{eqnarray}

The dust fluence on the VEGA 2 spacecraft can be modeled analogous
to the fluence on Stardust. At closest approach, VEGA 2 was also far
inside the envelope of the measured dust particles, therefore, the
total fluence was dominated by direct particles.

Because of the dependency of the integrand in equation (\ref{nproj})
on the dust escape velocity $v_{e,i}$, which, in turn, depends on the
mass of the dust particle, the mass distribution of particles which
hit the VEGA 2 instruments does not coincide with the mass
distribution of particles at the nucleus' surface, i.e. the
distribution of $Z_{d,i}$. Therefore a correction has to be applied to
the VEGA 2 fluence before it can be used as mass distribution at the
nucleus' surface. For this purpose we compute the dust escape
velocities $v_{e,i,{\rm Halley}}$ for an analogous radially symmetric
P/Halley model (The model parameters are
$Q_{\rm H_2O}=6.8\cdot10^{29}\;{\rm s}^{-1}$ (see Krankowsky (1986)
\cite{krank},
$Q_{\rm CO}=0\;{\rm s}^{-1}$, $r_n=5\;{\rm km}$; all other
model parameters are not changed). The fluence of dust class $i$ is
inversely proportional to the dust escape velocity $v_{e,i,Halley}$
for a radially symmetric coma model (see equation (\ref{nprojrsym})).
Therefore, the product $n_{srf,i}=n_{VG2,i} \cdot v_{e,i,{\rm
Halley}}$ is proportional to the mass distribution at the comet's
surface and the dust-to-gas mass ratios of the dust classes can be
computed by
\begin{eqnarray}
\chi_i & = & \chi_{\rm aux} \frac{m_{d,i}n_{srf,i}} {\sum_j
m_{d,j}n_{srf,j}},
\end{eqnarray}
where $\chi_{\rm aux}$ is an auxiliary dust to gas ratio. 
$\chi_{\rm aux}$ would have the meaning of the ratio of dust mass to 
gas mass released per unit time from the comet's nucleus, if all
dust classes could be lifted from the comet's surface. Since there
are dust classes which can not be lifted from the comet's surface or
can only be lifted inside some region around the sub-solar point, 
no physical interpretation can be linked to this number and is
in fact only an auxiliary value needed during the computation. 

\section{Model of LIC Dust Traversing the Solar System}
The impact parameters of interstellar particles on the dust collector
and the CIDA target are determined by the impact di\-rec\-tion and
velocity of the particles, and also by the articulation of the
spacecraft during the cruise phase. Since both is non-trivial, we
separate the effects by discussing two different types of particles:
(1) the reference particles as defined by the Stardust mission plan
\cite{stardust97} which move on straight lines through the Solar
System in a direction derived from Ulysses and Galileo data, and (2)
larger particles coming from the same direction, but being deflected
due to solar gravity. For the reference particles the change in impact
parameters is caused only by the spacecraft's trajectory and the
articulation strategies defined in the mission plan for the cruise
phase. Larger particles are mainly interesting for the dust-collector,
because CIDA is unlikely to be hit by such a particle due to its small
target area and the low abundance of these particles. Furthermore,
larger particles can be more easily extracted from the aerogel, this
extraction might be difficult for particles with sizes of about a few
tenth of a micron. In the following we argue why the reference
particles are a good approximation of the most abundant interstellar
particles, which are to be measured with Stardust and why smaller
interstellar particles are not expected to reach the spacecraft.

In general, the dynamics of a charged, massive dust-particle in the
interplanetary environment can be described by
\begin{eqnarray}
\ddot{\vec{r}} & = & -\gamma \left(1 - \beta\right) \frac{M_\odot}
{\left| \vec{r}\right|^3} + \frac{Q}{m}\left(\vec{v}_{\rm swf} \times
\vec{B} \right).\label{eqn_om}
\end{eqnarray}
This equation of motion takes into account gravity, radiation
pressure, and the Lorentz force induced by the solar wind magnetic
field. $\gamma$ is the coupling constant of gravity, $M_\odot$ is the
solar mass and the parameter $\beta$ is defined as the ratio of the
magnitudes of radiation pressure force and gravity. $Q$ is the
equilibrium charge to which a dust-particle will be charged in the
interplanetary radiation and plasma environment, $\vec{v}_{\rm swf}$
is the velocity vector of the particle, measured in the frame of the
radially expanding solar wind, and $\vec{B}$ represents the solar wind
magnetic field. In equation (\ref{eqn_om}) drag forces like
Poynting-Robertson drag and solar wind drag are neglected, because
these drag forces are small and only have a long-term effect on
particles on bound orbits. The dynamical parameters $\beta$ and $Q/m$
depend on the particle size and can be represented as functions of
particle radius $a$, if we assume a spherical shape for the particles
\cite{gustafson94,mukai81}.

\subsection{Reference Particles\label{straightlines}}
\begin{figure}[htb]
\hspace{-2mm}\epsfbox{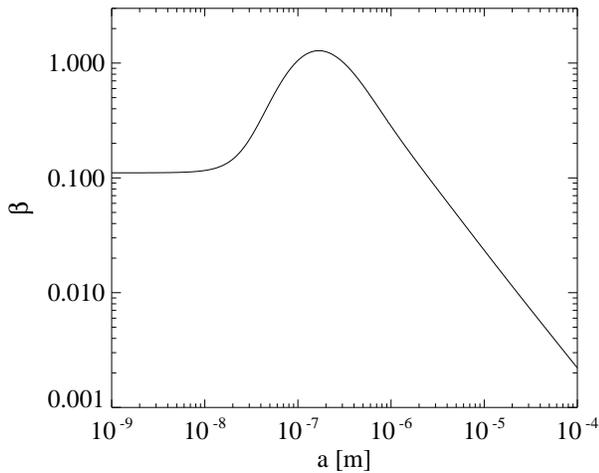}
\caption{\label{fig_beta}The parameter $\beta$ as a function of
particle radius $a$. Homogeneous, spherical particles with the optical
constants of astronomical silicates have been assumed (see
\protect\cite{gustafson94}).}
\end{figure}

Physically, the reference particles are implemented by setting $\beta=1$
and $Q/m=0\ {\rm C}\ {\rm kg}^{-1}$. In this case the solution to equation
(\ref{eqn_om}) are trajectories along straight lines,
i.e. $\dot{\vec{r}}=\vec{v}_{\infty,{\rm dust}}={\rm const.}$. Since
the Ulysses measurements indicate that interstellar dust and gas are
kinematically coupled, we set $\vec{v}_{\infty,{\rm
dust}}=\vec{v}_{\infty,{\rm gas}}$ using the parameters derived for
the gas by Witte et al. (1996) \cite{witte96}
Gustafson (1994) \cite{gustafson94} calculated $\beta$ as a function
of particle radius assuming a spherical shape and a composition
like the astronomical silicates defined by
Draine and Lee (1984) \cite{draine84} (see figure \ref{fig_beta}). Using
these calculations 
it was shown \cite{landgraf98b}, that for $57\%$ of all interstellar
particles measured by Ulysses and Galileo $|1-\beta|\leq 0.4$. If the
particles are measured at about $2\ {\rm AU}$, this deviation from the
reference behavior would change the impact angle by less than
$15^\circ$, which is the uncertainty in the determination of the
interstellar dust upstream direction from the Ulysses and Ga\-li\-leo
measurements. The impact velocity will deviate by less than $30\%$ for
these particles. Furthermore, the variation of impact direction and
velocity due to electro-magnetic interaction of interstellar
particles with the solar wind magnetic field ($Q/m \neq 0\ {\rm C}\
{\rm kg}^{-1}$) can be neglected if they have radii larger than $0.2\
{\rm
\mu m}$.

\subsection{Gravitational Particles\label{grav_grains}}
As shown in figure \ref{fig_beta}, $\beta \approx 1$ is not valid for
large particles. Large interstellar particles ($m>10^{-12}\ {\rm g}$)
are less abundant than smaller ($m<10^{-12}\ {\rm g}$) ones
\cite{landgraf96}, so they are more important for the dust collector
than for CIDA due to its larger surface area. If we neglect
electro-magnetic effects, the trajectories are hyperbolae and the
spatial density distribution of these particles can be calculated
analytically (see equation (\ref{nsngsol}) in Appendix A). Figure
\ref{fig_focus} shows this distribution as a grey-scale in the
ecliptic plane. In this simple model the density enhancement of large
particles due to gravitational focusing during the collection phase is
less than a factor of $2$.
\begin{figure}[htb]
\epsfbox{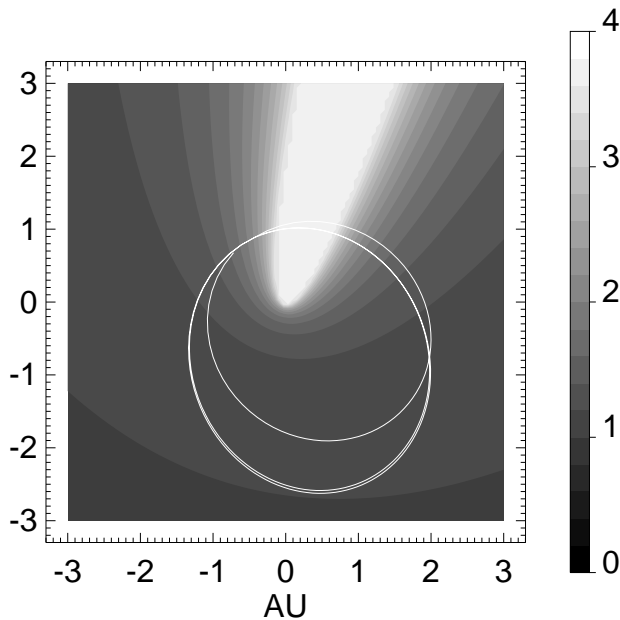}
\caption{ \label{fig_focus} The normalized density distribution of
gravitationally deflected ($\beta=0.1$) interstellar particles in the ecliptic plane
represented by a grey-scale. The normalization is defined to set the
spatial density at infinity equal to $1$. Stardust's trajectory is
indicated as the white solid line (compare to figure 1).}
\end{figure}
The enhancement in spatial density downstream the Sun is caused by
gravitational focusing. Since the down\-stream vector of the
interstellar dust flux has an angle of $\approx 5^\circ$ with respect
to the ecliptic plane, the actual focus does not lie in the ecliptic,
but a peripheral spatial density enhancement is present in the
ecliptic plane. The main effect of the dynamics of large particles
will be a higher impact velocity and a deviation in impact angle with
respect to the reference particles.

Particles with large $\beta$-values ($\beta > 1.4$) have diameters of
$\approx 0.45\ {\rm \mu m}$ which correspond to the maximum
wavelength of the solar spectrum. As we will argue in the next
section, we do not expect these particles to be abundant in the Solar
System during the Stardust mission due to their interaction with the
solar wind magnetic field.
\subsection{Electromagnetic Effects\label{emeffects}}
\begin{figure}[htb]
\epsfbox{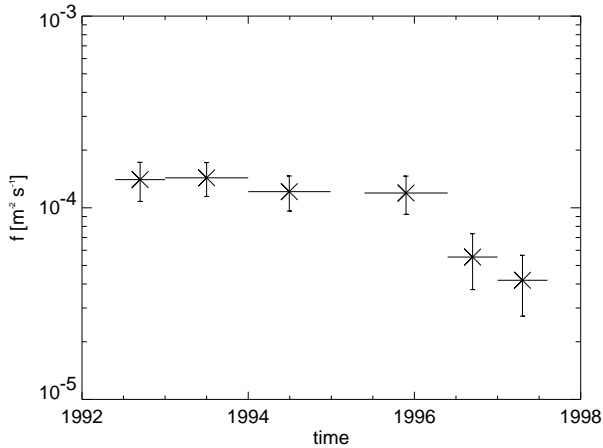}
\caption{\label{fig_ulsflux} Total flux $f$ of interstellar
particles measured by Ulysses as a function of time. The error-bars are
due to the fraction of particles which could not be clearly identified as
interstellar dust.}
\end{figure}

In section \ref{straightlines} we have argued that electromagnetic
effects can be neglected for the determination of the impact direction
and velocity of interstellar particles. But we have to take into
account these effects when considering the abundance of interstellar
particles in the Solar System since long-term effects might reduce or
enhance their spatial density. It was argued by Levy and Jokipii
(1976) \cite{levy76} that classical (very small) interstellar
particles are removed from the Solar System by Lorentz force. The
Ulysses and Galileo measurements show \cite{landgraf96} a depletion of
small particles (but still one order of magnitude above the detection
threshold). Models of the electromagnetic interaction of the charged
interstellar particles with the solar wind magnetic field
\cite{gruen94,gustafson96a,grogan96} predict a periodic focusing and
defocusing of the particles to the solar equator plane with the
$22$-years solar cycle. A new magnetic cycle starts at solar maximum
and the mean deflection effect is strongest when the magnetic field is
in the most ordered configuration during the solar minimum. The last
solar maximum in 1991 started a defocusing cycle. We have calculated
the total flux of interstellar particles on the Ulysses detector as a
function of time to check if the defocusing causes the total flux to
drop. Figure \ref{fig_ulsflux} shows that the total flux (in the
heliocentric inertial frame) of interstellar particles was constant
after Ulysses' fly-by of Jupiter in February 1992 (when Ulysses left
the ecliptic plane).
The gap in the data between beginning and mid of 1995 is caused by the
ecliptic crossing in March 1995 when distinction between interstellar
and interplanetary particles was difficult. After mid of 1996 the
measured total flux of interstellar particles drops by a factor of
$3$. We interpret this phenomenon as the effect of electromagnetic
defocusing. Unfortunately this can not be confirmed by Galileo data,
because Galileo is inside the Jovian system since the end of 1995,
therefore, Jupiter dust dominates the data and interstellar dust
impacts can not be identified \cite{krueger98a}. We assume that the
period of reduced total flux will last for half a solar cycle, which
is $11$ years, i.e. from mid of 1996 to mid of 2007. This period of
time covers the duration of the Stardust mission. Following this
estimate, we predict from the Ulysses data that the total dust flux on
the Stardust experiments will be reduced by a factor of $3$ compared
to the Ulysses and Galileo measurements.

For Ulysses and Galileo a total flux of $f_{\rm U/G} = 1.5\cdot
10^{-4}$ ${\rm m}^{-2}\ {\rm s}^{-1}$ has been determined
\cite{baguhl95b}, so we predict $f_{\rm SD}=5\cdot 10^{-5}\ {\rm
m}^{-2}\ {\rm s}^{-1}$.

\section{Results}
\subsection{Cometary Dust Measurements}

In this section we apply the coma model developed in section
\ref{commod} to quantitatively predict the flux and the fluence of
cometary dust particles on the Stardust spacecraft during the P/Wild 2
encounter.  The results we obtain for an axially symmetric model are
compared with the results we get if we use a radially symmetric
model. We can compare our results for Stardust at P/Wild 2 with the
fluence measured by the VEGA spacecraft inside the coma of P/Halley by
scaling the predicted fluence with the difference in brightness,
encounter distance, and phase angle.

\paragraph{Fluence as a function of mass}
We approximate the trajectory of Stardust as a straight line passing
the coma with a phase angle of $70^\circ$ with respect to the Sun
direction. Stardust's closest approach to the nucleus is above the
comet's day side. The fluence of dust particles of a given dust class
$i$ during the whole trajectory is equal to the column density along
the spacecraft's trajectory which is given by equation
(\ref{nproj}). As the fluence is inversely proportional to the
distance of closest approach of the spacecraft to the nucleus $\rho$,
our results can easily be scaled to other closest approach
distances. The numerical values of the fluence for a trajectory above
the sub-solar point is given in table \ref{cfltab}.

\paragraph {Flux and Fluence as a Function of Spacecraft Position}

The fluence as a function of time from closest approach is calculated
by the time integral, or, equivalently, by the integral over the path
$s$ along the spacecraft's trajectory, of the dust flux on the
spacecraft as given in equation (\ref{fsc}).
\begin{eqnarray}
F_{SC,i}(t) & = & \int_{-\infty}^{V_{SC}t} f_{SC}\;dt =
\int_{-\infty}^{s(t)}n_{d,i}(\vec r(s))\;ds
\end{eqnarray}
\begin{table}
\caption{\label{cfltab} Fluence of dust particles on the spacecraft
for the radial and axially symmetric model for a closest approach
distance of $\rho=100\;{\rm km}$. For different distances
$\rho^\prime$ of closest approach the fluence can be scaled by
$\rho/\rho^\prime$. In the axially symmetric model we assume that the
spacecraft's trajectory crosses the sub-solar point.}
\begin{tabular}{lllll}
\hline
Dust & particle & particle && \\
class & mass & radius &
\multicolumn{2}{c}{fluence $[{\rm m}^{-2}]$}
\\ 
$i$ & $m_{d,i}\ [{\rm kg}]$ & $s_{i}\ [{\rm m}]$ & radial-symm. & 
axial-symm.\\ 
\hline 
$1$&$
10^{-20}$&$ 1.34\cdot 10^{-8}$&$ 5.86\cdot 10^{+07}$&$
1.11\cdot 10^{+08}$\\ $2$&$ 10^{-19}$&$ 2.88\cdot 10^{-8}$&$
3.86\cdot 10^{+07}$&$ 7.05\cdot 10^{+07}$\\ $3$&$ 10^{-18}$&$
6.20\cdot 10^{-8}$&$ 2.52\cdot 10^{+07}$&$ 4.46\cdot 10^{+07}$\\ $4$&$
10^{-17}$&$ 1.34\cdot 10^{-7}$&$ 1.60\cdot 10^{+07}$&$
2.76\cdot 10^{+07}$\\ $5$&$ 10^{-16}$&$ 2.88\cdot 10^{-7}$&$
9.67\cdot 10^{+06}$&$ 1.64\cdot 10^{+07}$\\ $6$&$ 10^{-15}$&$
6.20\cdot 10^{-7}$&$ 5.39\cdot 10^{+06}$&$ 8.98\cdot 10^{+06}$\\ $7$&$
10^{-14}$&$ 1.34\cdot 10^{-6}$&$ 2.58\cdot 10^{+06}$&$
4.26\cdot 10^{+06}$\\ $8$&$ 10^{-13}$&$ 2.88\cdot 10^{-6}$&$
9.37\cdot 10^{+05}$&$ 1.53\cdot 10^{+06}$\\ $9$&$ 10^{-12}$&$
6.20\cdot 10^{-6}$&$ 2.32\cdot 10^{+05}$&$ 3.75\cdot 10^{+05}$\\ $10$&$
10^{-11}$&$ 1.34\cdot 10^{-5}$&$ 4.11\cdot 10^{+04}$&$
6.62\cdot 10^{+04}$\\ $11$&$ 10^{-10}$&$ 2.88\cdot 10^{-5}$&$
6.02\cdot 10^{+03}$&$ 9.64\cdot 10^{+03}$\\ $12$&$ 10^{-09}$&$
6.20\cdot 10^{-5}$&$ 8.11\cdot 10^{+02}$&$ 1.29\cdot 10^{+03}$\\ $13$&$
10^{-08}$&$ 1.34\cdot 10^{-4}$&$ 1.06\cdot 10^{+02}$&$
1.67\cdot 10^{+02}$\\ $14$&$ 10^{-07}$&$ 2.88\cdot 10^{-4}$&$
1.37\cdot 10^{+01}$&$ 2.14\cdot 10^{+01}$\\ $15$&$ 10^{-06}$&$
6.20\cdot 10^{-4}$&$ 1.78\cdot 10^{+00}$&$ 2.72\cdot 10^{+00}$\\ $16$&$
10^{-05}$&$ 1.34\cdot 10^{-3}$&$ 2.40\cdot 10^{-01}$&$
3.44\cdot 10^{-01}$\\ $17$&$ 10^{-04}$&$ 2.88\cdot 10^{-3}$&$
3.55\cdot 10^{-02}$&$ 4.34\cdot 10^{-02}$\\ $18$&$ 10^{-03}$&$
6.20\cdot 10^{-3}$&$ 6.31\cdot 10^{-03}$&$ 5.37\cdot 10^{-03}$\\ $19$&$
10^{-02}$&$ 1.34\cdot 10^{-2}$&$ 0.00\cdot 10^{+00}$&$
5.71\cdot 10^{-04}$\\ $20$&$ 10^{-01}$&$ 2.88\cdot 10^{-2}$&$
0.00\cdot 10^{+00}$&$ 1.15\cdot 10^{-05}$\\ \hline
\end{tabular}
\end{table}

For a radially symmetric model an analytical expression is given by:
\begin{eqnarray}\label{eqn_fsci_radsymm}
F_{SC,i} & = & \int_{-\infty}^{V_{SC}t} \frac{Z_{e,i}} {V_{e,i}
\rho^2+s^2}\;ds\nonumber \\
& = & \frac{Z_{e,i}} {V_{e,i}\rho} \cdot \left(\xi +
\frac{\pi} {2}\right),
\end{eqnarray}
where $\xi = \arctan(V_{SC}t / \rho)$ is the angle of the spacecraft
position to the point of closest approach. From equation
(\ref{eqn_fsci_radsymm}) we see that $F_{SC,i}$ is a linear function
of $\xi$ in the radially symmetric model.

\begin{figure}[htb]
\epsfxsize=.5\hsize
\epsfbox{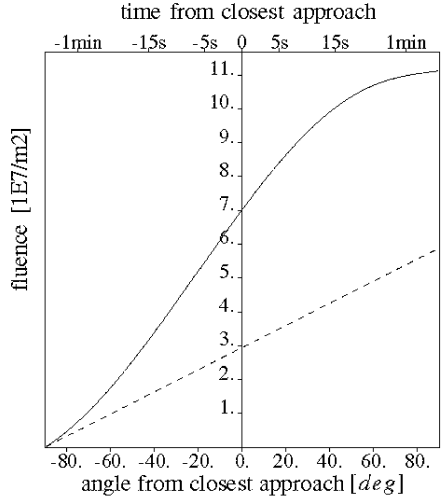}
\caption{\label{uflu}
Total fluence on Stardust versus the angle $\xi$ from the location of
closest approach for the radially symmetric (dashed line) and axially
symmetric model (solid line). The total fluence calculated by the
axially symmetric model deviates significantly from the result of the
radially symmetric calculation.}
\end{figure}

We show the calculated total dust fluence on Stardust in figure
\ref{uflu}. 
\begin{figure}[htb]
\epsfxsize=.5\hsize
\epsfbox{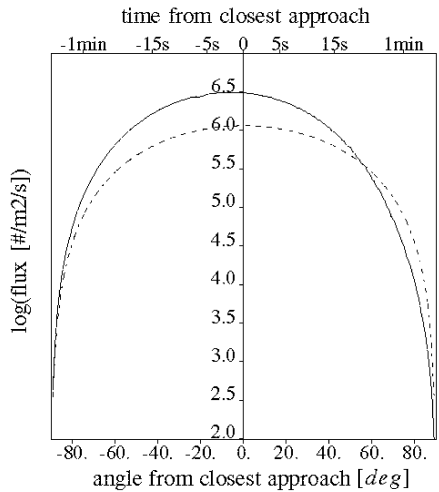}
\caption{\label{uflx}
Total dust flux versus the angle $\xi$ between Stardust's location
on the trajectory and the location of closest approach P/Wild 2. The
total flux of the axis and radially symmetric model are represented by
the solid and dashed lines, respectively. It can be seen that the
total flux predicted by the radially symmetric model is symmetric with
respect to closest approach and reaches its maximum at that
point.}
\end{figure}
In the axially symmetric model the total fluence is approximately a
linear function of $\xi$ for $-40^\circ \leq \xi \leq 40^\circ$ ,
like it would be for the radially symmetric model. The radially
symmetric model predicts a too low dust flux during closest approach,
because it does not take into account the enhanced activity on the day
side. Therefore, the slope of the total fluence is flatter between
$\xi=-60^\circ$ and $\xi=60^\circ$ in the radially symmetric
case. For $\xi > 60^\circ$ Stardust enters the region above the
terminator and consequently the total flux (and therefore the slope of
the total fluence) drops below the value predicted by the radially
symmetric model. We conclude that, for predicting the total dust
fluence on Stardust during closest approach, one must take into
account that Stardust approaches the comet from the day side and
passes the terminator shortly after closest approach. The different
results from the radially and axially symmmetric calculations show
that the total fluence is strongly affected by the activity
distribution on the surface of the nucleus. Unfortunately, we do not
have any information about the activity distribution on P/Wild 2, so
we have to rely on the axially symmetric coma model as a simple yet
complete physical model of the inner coma for which we can constrain
the parameters by measurements.

In figure \ref{uflx} we show the total particle flux versus
$\xi$. In the radially symmetric model the closest approach coincides
with the point of highest total dust flux. Using the axially symmetric model,
we find that the maximum total flux is reached $3\;{\rm s}$ before closest
approach and that the total flux stays nearly constant until after closest
approach. This is because the spacecraft enters into a region of lower
activity after it has passed the sub-solar point but the decrease
of cometary activity is apparently compensated by the still decreasing
distance from the nucleus.

\paragraph{Reproduction of the VEGA and Giotto measurements at P/Halley}
To validate our prediction of the fluence on the Stardust instruments,
we scale our model to the conditions at P/Halley during the fly by of
the Giotto and VEGA spacecraft. We stress that, although the VEGA 2
data were used to derive a mass distribution of our coma model, the
comparison of the fluence is in fact a check of the model. This is
because the VEGA 2 data were only used to derive the relative
abundance of particles in different dust classes, but we derive the
total amount of the dust activity from the value of
$Af\rho$ that has been determined from ground based observations.

\begin{figure}[htb]
\epsfxsize=.5\hsize
\epsfbox{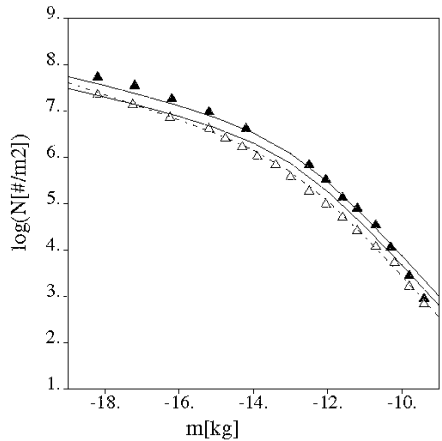}
\caption{\label{whcmp}
Fluence of dust particles on the VEGA spacecraft. Filled and outlined
triangles represent VEGA 1 and VEGA 2 data, respectively. The fit to
the VEGA 2 data by Divine and Newburn (1987) \protect\cite{div2} is
shown as the dashed
line and the solid lines represent the scaled fluences as predicted by
the axially symmetric (upper solid curve) and radially symmetric model
(lower solid curve)}
\end{figure}

The fluence on the Stardust and VEGA spacecraft is proportional to
the value of $Af\rho$ of the coma, inversely proportional to the
distance $\rho$ of closest approach, and the dust phase function
during the observation. Thus, the fluence which was computed for the
Stardust spacecraft can be scaled to the fluence measured on board the
VEGA spacecraft by multiplying the fluence with the following three
ratios:
\begin{equation}
\begin{array}{ccccc}
\multicolumn{5}{l}{\mbox{closest approach distance ratio:}} \\
\frac{\mbox{$\rho_{\mbox{\tiny Stardust}}$}} {\mbox{$\rho_{\mbox{\tiny
VEGA}}$}}
&\approx &\frac{100\;{\rm km}}{8000\;{\rm km}} &=& \frac{1} {80}
\\
\\
\multicolumn{5}{l}{\mbox{$Af\rho$-ratio:}}\\
\frac{\mbox{$Af\rho_{\mbox{\tiny Halley}}$}}
{\mbox{$Af\rho_{\mbox{\tiny Wild 2}}$}}& \approx & 
\frac{219\;{\rm m}}{4.27\;{\rm m}} & \approx & 51.3
\\
\\
\multicolumn{5}{l}{\mbox{dust phase function ratio:}}\\
\frac{\mbox{$j(\alpha_{\mbox{\tiny Wild 2}}\approx 30^\circ)$}}
{\mbox{$j \left(\alpha_{\mbox{\tiny Halley}}
\approx 60^\circ\right)$}} &
\approx &\frac{0.04} {0.032}& = & 1.25
\end{array}
\end{equation}
The $Af\rho$ value of P/Halley during the VEGA encounter was taken
from Schleicher et al. (1998) \cite{afrhalley}. This leads to a total
scaling factor
of $51.3\cdot 1.25 / 80\approx 0.8$.  Hence the fluence which we
predict for Stardust at P/Wild 2 is compatible with the fluence
measured on the VEGA spacecraft at P/Halley. In figure \ref{whcmp} we
show the data of the fluence measured by VEGA together with the our
prediction of the fluence on Stardust scaled by $0.8$.
The good agreement of
the number of dust particles collected on the VEGA spacecraft with
the scaled Stardust fluence validates our procedure to derive the total dust
activity of the comet from the observations and it shows that the
parameter set we use in our model is consistent.

Unfortunately, because of the loss of the Giotto data near closest
approach due to large particle impacts, we can not perform any
quantitative comparison of the Giotto fluence to our model. However,
we can estimate the total mass fluence on Giotto on the basis of the
well known deceleration of $\Delta V=23\;{\rm cm}\;{\rm s}^{-1}$
\cite{eden} of Giotto during the encounter. As for the considered
mass distribution, the total fluence is dominated by large particles
and the momentum enhancement factor is believed to
be rather low for large particles \cite{eden},
the total mass fluence is estimated by assuming inelastic dust particle
impacts on Giotto: With the Giotto mass of $M_{\rm Giotto}=573.7\;{\rm
kg}$ and the velocity of $V_{\rm Giotto}=68.37\;{\rm km}\;{\rm
s}^{-1}$ during encounter, the total mass which hit Giotto can be
estimated to be $m_{\rm Giotto}=M_{\rm Giotto}/$ $V_{\rm Giotto}\cdot
\Delta V\approx 2\;{\rm g}$. Using the area of the
Giotto dust shield of $A=2.64\;{\rm m}^2$, we determine the total mass
fluence on Giotto to be $0.76\;{\rm g}\;{\rm m}^{-2}$. The total mass
fluence on the Stardust at P/Wild 2 can also be scaled to the total
mass fluence on Giotto at P/Halley by taking into account that the
closest approach distance of Giotto was approximatively $600\;{\rm
km}$. We find the values $0.038\;{\rm g}$ and $0.064\;{\rm g}$ for the
radially symmetric and axially symmetric model, respectively. Thus, the
extrapolation of the mass distribution which was fitted by Divine and
Newburn (1987) \cite{div2} 
to the fluence on the VEGA 2 spacecraft to larger masses leads to an
underestimation of the total mass fluence by about one order of magnitude
when compared to the total mass fluence derived from the final impact
on Giotto. We discuss this discrepancy and its significance in section
\ref{discussion}.

\subsection{Interstellar Dust Measurements}
We predict the measurements of interstellar dust during the
interplanetary cruise phase by both main instruments on-board
Stardust, CIDA and the aerogel collector. Unlike in the case of the
measurements at P/Wild 2, the geometry of the measurements of both
instruments is different, because they are used at different parts of
the interplanetary trajectory. Therefore, we discuss the predictions
for both instruments sepa\-rate\-ly.

\subparagraph{CIDA}
To comprehend the predictions we make for CIDA one has to understand
the three attitude strategies which have been defined by the mission
plan \cite{stardust97}. Since CIDA has a fixed position on Stardust,
its pointing is determined by the spacecraft attitude. The overall
strategy should be to choose the attitude in a way that a maximum
number of interstellar particles are detected assuming that they all
behave like the reference particles. But the more one optimizes the
attitude, the more complicated the spacecraft operations will become
during the cruise phase. Thus, a trade-off has to be found between
maximum number of detected particles and operation complexity.

\begin{figure}[htb]
\epsfbox{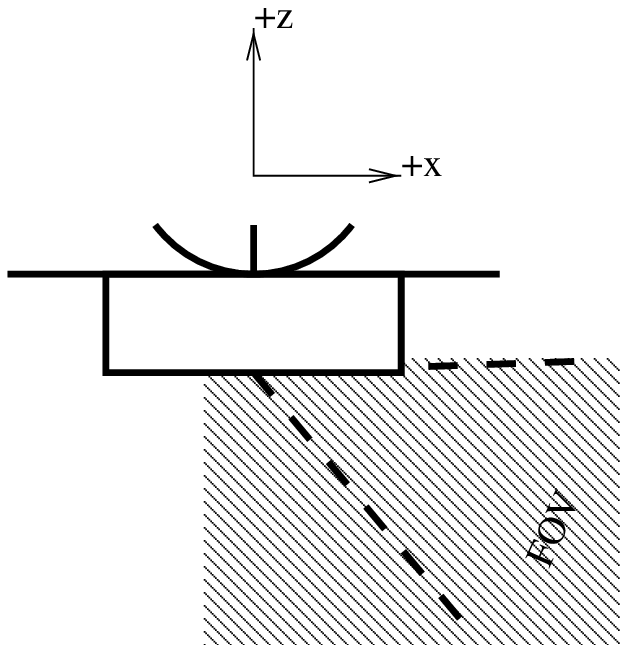}
\caption{\label{fig_cidaconfig} Configuration of the CIDA field of
view (FOV) with respect to Stardust. The $+z$-direction is defined by
the pointing of the high gain antenna and the fixed solar panels,
which should always be pointed roughly towards the Sun.}
\end{figure}

We now explain the three different attitude strategies defined in the
mission plan. Figure \ref{fig_cidaconfig} shows a simplified sketch of
Stardust containing the high gain antenna defining the $+z$-direction
, the solar panels, which lie in the $x$-$y$-plane and the CIDA field
of view (FOV) in the $x$-$z$-plane.
In this plane the angle of FOV is $52^\circ$. Strategy \#1 fixes the $+x$-axis
to the upstream-direction of reference particles. This means that the
$+z$ axis, i.e. the solar panels and the high gain antenna, are not
always pointing to the Sun. The mission plan restricts this off-pointing to
$45^\circ$ due to power concerns. If more than $45^\circ$ off-pointing
is required to keep the $+x$-axis pointing upstream, the spacecraft
is to be turned to full Sun-pointing again, which means that the
upstream-direction is turned out of the FOV. Strategy \#2 is the most
complex one. If pointing of the $+x$-axis towards the upstream
direction is possible without more than $45^\circ$ off-pointing of the
$+z$-axis with respect to the Sun, strategy \#2 is identical to
strategy \#1. But if this configuration is not possible any more,
strategy \#2 keeps the $45^\circ$ off-pointing as long
as the upstream-direction lies inside the FOV. If even this is
impossible, the spacecraft should return to Sun-pointing again. The
most simple strategy is strategy \#3. It fixes Sun-pointing of the
$+z$-axis all the time, so CIDA will only collect interstellar particles
when the upstream direction of the flux is occasionally in the
FOV. Table \ref{tab_strategies} summarizes the three attitude
strategies.
\renewcommand{\arraystretch}{2}
\begin{table}
\caption{\label{tab_strategies} Attitude strategies for the
Stardust mission during the cruise phase as given by the mission plan.}
\begin{tabular}{ll}
\hline
strategy & description \\
\hline
\# 1 & 
\begin{minipage}[t]{6cm}
$+x$-axis points upstream. Maximum Sun off-pointing is $45^\circ$.
\end{minipage}\\
\# 2 &
\begin{minipage}[t]{6cm}
upstream-direction is kept in the FOV. Maximum Sun off-pointing is
$45^\circ$. 
\end{minipage}\\
\# 3 & 
\begin{minipage}[t]{6cm}
$+z$-axis always points towards the Sun.
\end{minipage}\\
\hline
\end{tabular}
\end{table}
\renewcommand{\arraystretch}{1}

\begin{figure*}[htb]
\epsfxsize=\hsize\epsfbox{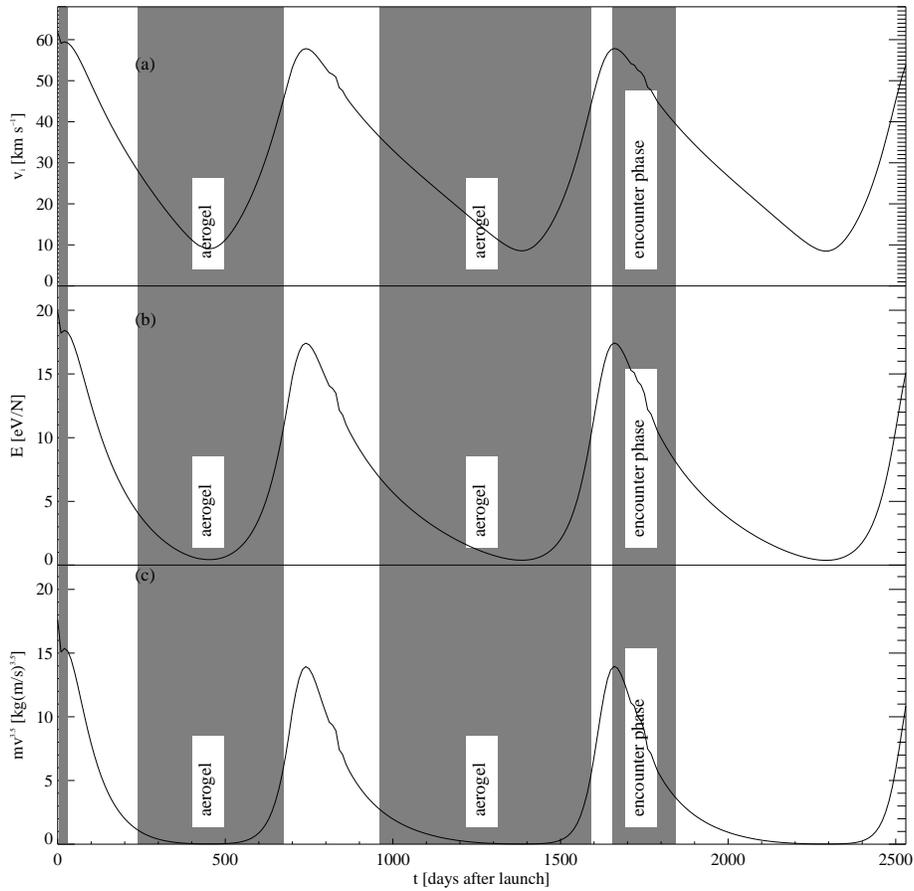}
\caption{\label{fig_imp_ref} The values of $v$, $E=mv^2/2$, and $m
v^{3.5}$ as a function of time, where $m$ is the mass and $v$ the
velocity of the impacting reference particle. The shaded areas labeled
``Arogel'' indicate the phases of deployment of the Aerogel collector
and ``encounter Phases'' labels the phase of preparations of the comet
encounter. During these phases, CIDA is not taking data.}
\end{figure*}

Since the reference particles have a constant velocity of $26\;{\rm
km}\;{\rm s}^{-1}$, the impact velocity of the reference particles as
a function of time is determined by the spacecraft velocity. The
impact velocity, the impact energy, and the quantity $m v_i^{3.5}$,
which is proportional to the impact charge for impact plasma detectors
\cite{gruen95a}, where $m$ is the mass and $v_i$ is the impact velocity
of the particle, are shown in figure
\ref{fig_imp_ref}.

\begin{figure*}[htb]
\epsfxsize=\hsize\epsfbox{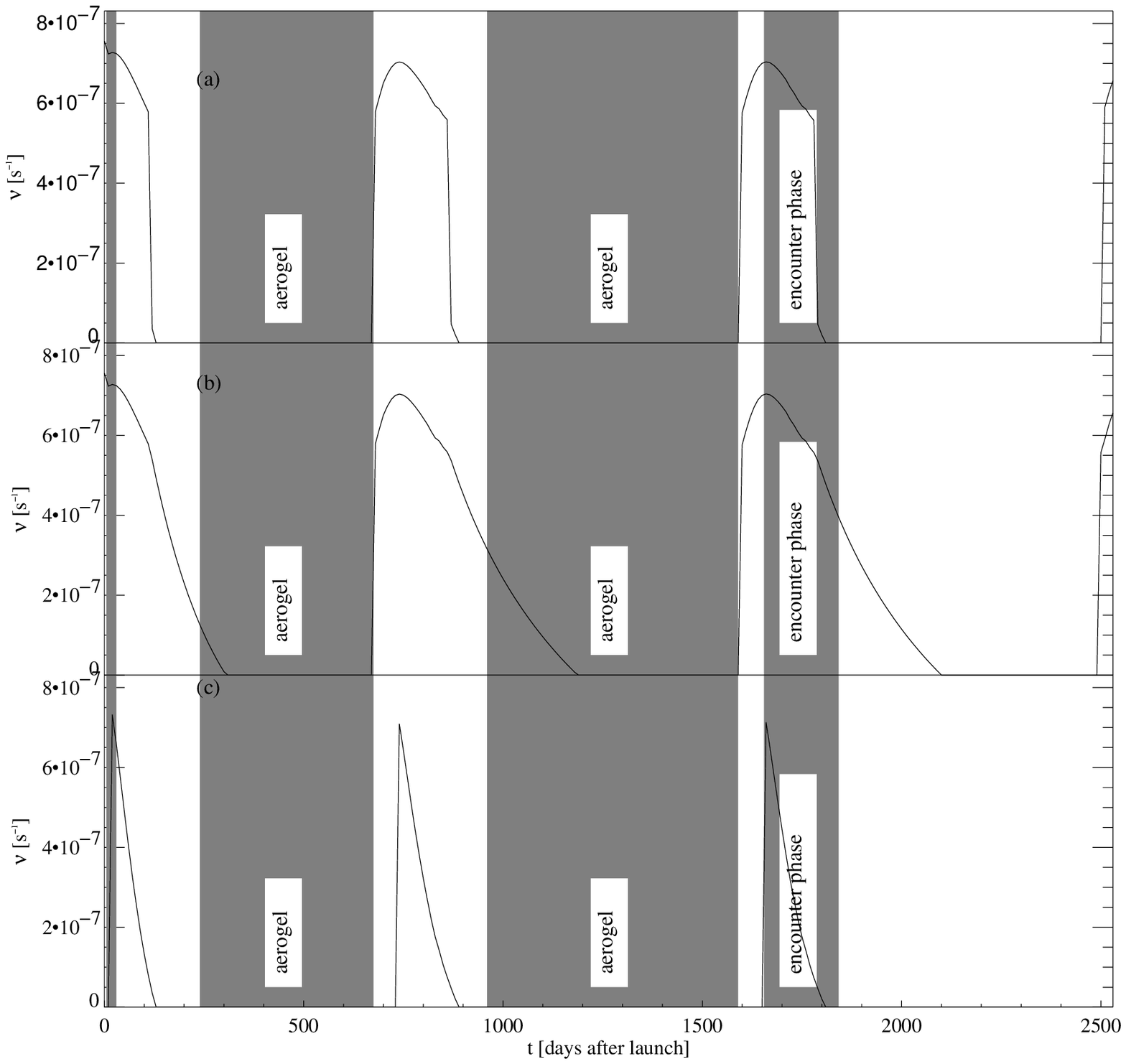}
\caption{\label{fig_rate} Impact rate on the CIDA target as a
function of time. Prediction in panel (a) assumes strategy \#1, in
panel (b) strategy \#2, and in panel (c) strategy \#3. Mission phases
during which CIDA is not allowed to be turned on due to other
activities are indicated by the shaded areas (compare figure
\ref{fig_imp_ref}).}
\end{figure*}

We predict the impact rate $\nu$ of reference particles on the CIDA
target. We can estimate $\nu$ by the given effective sensitive target
area of $A_{\rm CIDA} = 80\ {\rm cm}^2\cdot \cos 40^\circ = 60\;{\rm
cm}^2$
\cite{jochenpers}, the flux $f_{\rm SD}=5\cdot 10^{-5}\;{\rm
m}^{-2}\;{\rm s}^{1}$ (see section
\ref{emeffects}) in the inertial heliocentric frame, and the enhancement due
to the upstream motion of Stardust. We get
\begin{eqnarray}
\nu_{\rm max} & = & A_{\rm CIDA} f_{\rm SD} \left(\frac{v_{\rm
rel,max}} {v_{\rm dust,ecl}} \right) \approx 7.0 \cdot 10^{-7}\ {\rm
s}^{-1},
\end{eqnarray}
where $v_{\rm rel,max}=60\ {\rm km}\ {\rm s}^{-1}$ is the maximum
relative velocity of Stardust to the stream of reference particles (see
above) and $v_{\rm dust,ecl}=26\ {\rm km}\ {\rm s}^{-1}$ is the
velocity of the reference particles in the inertial ecliptic
frame. Figure \ref{fig_rate} shows the predicted impact rate on the
CIDA target as a function of time for all three strategies explained
above.

\begin{figure*}[htb]
\epsfxsize=\hsize\epsfbox{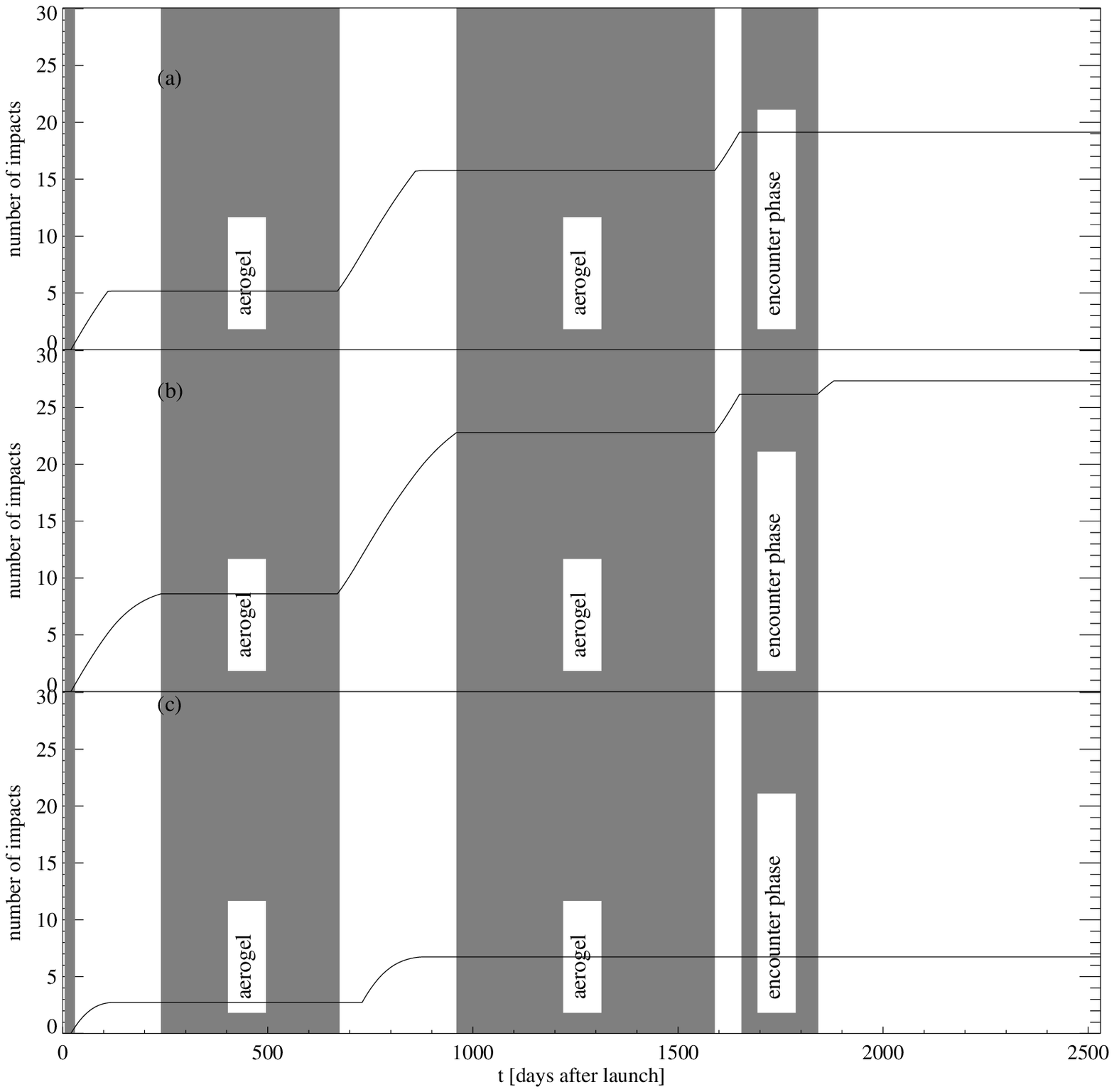}
\caption{\label{fig_fluence_ref} Total fluence on the CIDA target as a
function of time. Prediction in panel (a) assumes strategy \#1, in
panel (b) strategy \#2, and in panel (c) strategy \#3. Mission phases
during which CIDA is not allowed to be turned on due to other
activities are indicated by the shaded areas (compare figure
\ref{fig_imp_ref}).}
\end{figure*}

The total number of reference particles that CIDA will detect is given by
the accumulated impact rate over time. We show the prediction in figure
\ref{fig_fluence_ref}.

In summary, we predict that, assuming attitude strategy \#2, CIDA will
detect about $25$ particles which behave very much like reference
particles and hit the target with velocities between $30\ {\rm km}\
{\rm s}^{-1}$ and $60\ {\rm km}\ {\rm s}^{-1}$ and energies between
$18\;{\rm eV}\;{\rm N}^{-1}$ and $4\;{\rm eV}\;{\rm N}^{-1}$ (${\rm
eV}\;{\rm N}^{-1}$ = electronvolt per nucleon). For a summary of the
prediction see table \ref{tab_ip}, this table also contains the number
of interplanetary dust particles as determined from the
five-population model of interplanetary dust by Staubach et al. (1997)
\cite{staubach97} ``contaminating'' the measurement.
\begin{table}[ht]
\caption{\label{tab_ip} Number of interstellar and interplanetary
particles predicted to be measured by Stardust during the cruise. For
the collector we give the numbers for the front (which is exposed to
the comet) and the back side (which is exposed to the interstellar
upstream direction) seperately.}
\begin{tabular}{rr|rr}
\hline
& & \multicolumn{2}{c}{collector}\\
& CIDA\footnotemark[1]
& front & back \\
\hline
interstellar & $25$ & $0$ & $120$\\
interplanetary & $5$ & $25$ & $20$\\
\hline
\multicolumn{4}{l}{\footnotesize $^1$ Attitude strategy \#2, see table
\ref{tab_strategies}}
\end{tabular}
\end{table}
\clearpage

\subparagraph{The Dust Collector}
\begin{figure*}[htb]
\epsfxsize=\hsize\epsfbox{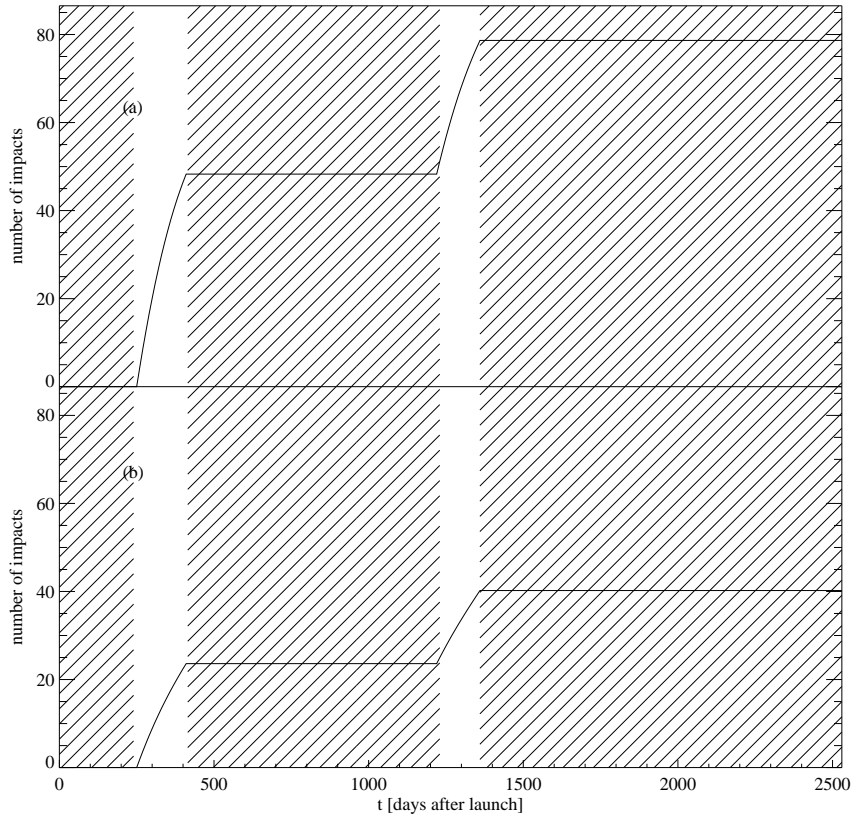}
\caption{\label{fig_coll_fluence} Total fluence on the aerogel detector
of reference particles (a) and gravitationally deflected particles (b) as a function of
time. Aerogel collection is not possible in the hatched periods.}
\end{figure*}
\begin{figure*}[htb]
\epsfxsize=\hsize\epsfbox{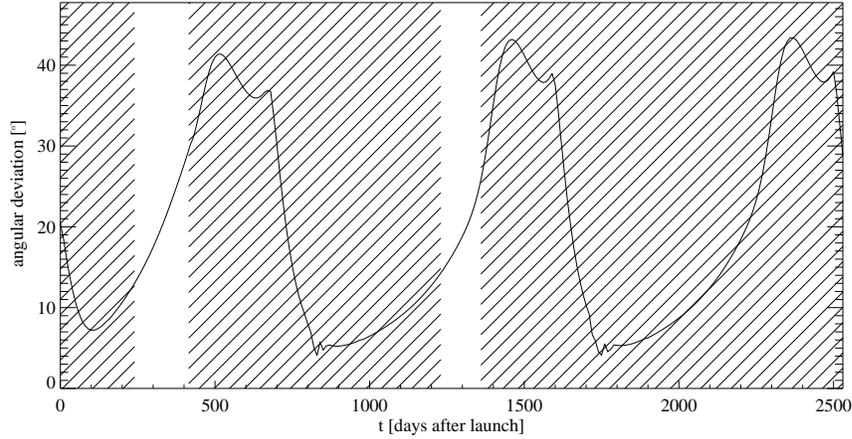}
\caption{\label{fig_dev_angle} Deviation of the impact angle of
gravitationally deflected particles from the impact angle of reference particles, which
should be close to $0^\circ$ as a function of time. Aerogel collection
is not possible in the hatched regions.}
\end{figure*}
\begin{figure*}[htb]
\epsfxsize=\hsize\epsfbox{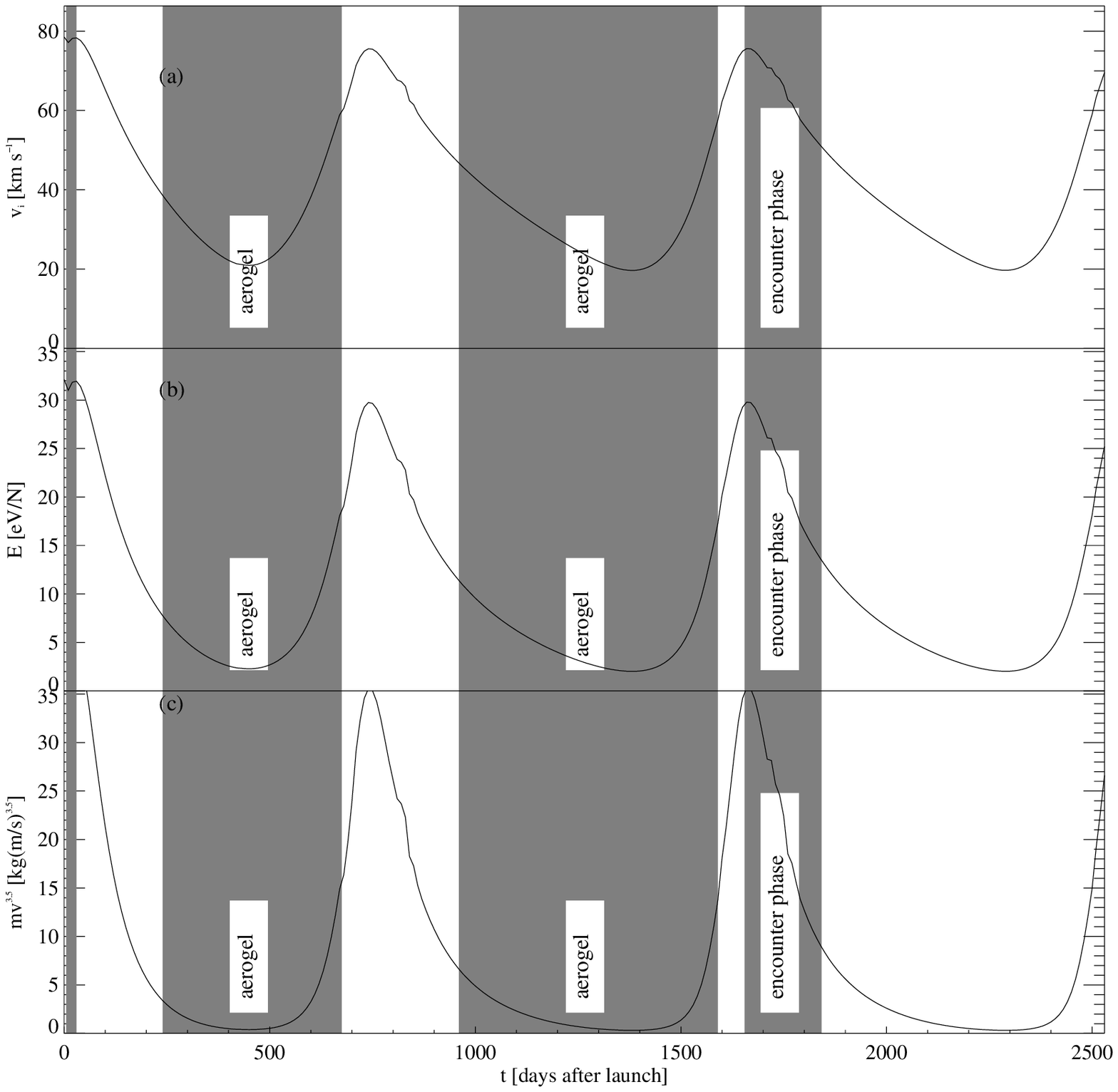}
\caption{\label{fig_imp_grav} Values of $v$, $E=mv^2/2$, and $m
v^{3.5}$ as a function of time, where $m$ is the mass and $v$ the
velocity of the impacting gravitationally deflected particle (compare figure
\ref{fig_imp_ref}).}
\end{figure*}
For the dust collector larger particles are more interesting, because
their trajectories can more easily be measured and they can be
extracted from the aerogel. Since the collector is turned to maintain
a pointing to the upstream direction of the reference particles, we
expect a spread in impact direction due to dynamical effects. For
larger particles the dynamics should be dominated by gravity and we
assume $\beta=0.1$ (see section \ref{grav_grains}). The
Mie-calculations
\cite{gustafson94} give a radius of $1\ {\rm \mu m}$ for these particles,
which translates to a mass of $1\cdot 10^{-11}\ {\rm g}$ assuming a
density of $2.5\ {\rm g}\ {\rm cm}^{-3}$. In the Ulysses and Galileo
data particles heavier than $10^{-11}\ {\rm g}$ make up $\approx 10\%$ of
the total number \cite{landgraf96}, but they are not depleted due to
electro-magnetic effects (see section \ref{emeffects}). So we assume a
flux of $1.5\cdot 10^{-5}\ {\rm m}^{-2}\ {\rm s}^{-1}$ of these large
particles. We show the total fluence of interstellar particles on the
collector for both, reference and gravitationally deflected
particles, in figure \ref{fig_coll_fluence}.

Due to particle dynamics, tracks of larger particles should show a direction
distribution. We show the deviation of impact angle as a function of
time in figure \ref{fig_dev_angle}.

For the collector it is furthermore interesting how fast
gravitationally deflected
particles impact into the aerogel. Since the particles are accelerated
towards the Sun we expect their impact velocities to be higher than
the impact velocities of the reference particles. In figure
\ref{fig_imp_grav} we show the same plot as in figure
\ref{fig_imp_ref} for gravitationally deflected particles.

To summarize our prediction for the aerogel collector, we give the
total number particles collected during the $290$ days of exposure time
\cite{stardust97} to be $120$. $80$ of these particles are small, like
reference particles. The remaining $40$ are large particles, the impact
tracks of which have angular deviations between $10^\circ$ and
$30^\circ$. For the large particles we expect impact velocities on the
aerogel between $20\ {\rm km}\ {\rm s}^{-1}$ and $40\ {\rm km}\ {\rm
s}^{-1}$.
\clearpage

\section{Discussion\label{discussion}}
We think that the Stardust mission to comet P/Wild 2 will enhance our
understanding of the environment and the properties of comets by a
large amount. According to our prediction, Stardust is able to fulfill
its primary goal, the in situ measurement and collection of cometary
dust particles. Concerning the secondary goal, the in situ detection
and collection of interstellar dust particles from the local
interstellar cloud, we find that the total number of
particles we expect to be measured by Stardust is quite small
considering the seven year duration of the mission. But having
material from beyond the Solar System in a laboratory on the ground
would be a big achievement.

We have determined the fluence on the Stardust spacecraft using
a coma model. In our model the dust activity was determined using
observations of P/Wild 2 at $1.7\;{\rm AU}$ heliocentric distance. Our
model can reproduce the measurements of the VEGA spacecraft at
P/Halley when scaling the model to the geometry of this measurement. The
total mass fluence on Giotto, deduced from the fact that the spacecraft
was hit by a very large particle, is underestimated by our model by
one order of magnitude. Of course, a one-particle hit has not a very
high statistic significance and we can hypothesize that this one hit was
a low probability event. The possibility that the mass distribution
used in our model has a deficit of large dust particles, as suggested
by the DIDSY data \cite{fulle}, exists, but
simply adding large particles to the mass distribution would either
contradict ground based observations of the total brightness of P/Wild
2 during its in-bound part of its orbit at $1.7\;{\rm AU}$, or the
dust fluence measured by VEGA at P/Halley. The real
mass distribution of dust particles at the nucleus is still unknown
and the reader may choose a different mass distribution and put it into
our calculation to calculate the expected flux on Stardust.

Stardust approaches the comet from the sunlit side and therefore more
dust particle impacts are expected before than after
closest approach. Thus, a radially symmetric model is not
sufficient to calculate the dust fluence on CIDA and the aerogel
collector. As a result of the axially symmetric model, radial features
on the night side of the nucleus appear (see figure
\ref{dstiso}). Unfortunately, we can not prove or disprove the
existence of these features by the Stardust measurements, because
Stardust is passing by the day side of P/Wild 2. The prove or disprove
of the existence of such features remains a task for future ground
based observations and space missions.

On the basis of the Galileo and Ulys\-ses measurements we predict that
during the interplanetary cruise phase $25$ mainly small
($m\approx 10^{-12}\ {\rm g}$) particles will hit the CIDA target. As the
Ulysses data indicate, Stardust's timing with the solar magnetic cycle
is unfortunate (see section \ref{emeffects}). To confirm that the
decrease in the total flux of interstellar particles measured by
Ulysses is due to their interaction with the solar wind magnetic
field, modeling of this interaction is under way
\cite{landgraf98b}. Even if the total number of interstellar
particles detected by CIDA could be low, the scientific significance
of the data, in any case, is high because the composition of LIC dust
is to be directly measured for the first time by CIDA. The
reliability of this measurement depends very much on the ability to
identify interstellar impacts clearly. Interplanetary particles are a
possible source of ``contamination'' for CIDA's measurements. Table
\ref{tab_ip} shows that 5 interplanetary dust particles with masses
greater than $10^{-14}\;{\rm g}$ are expected to hit the CIDA target
during the cruise phase. For the determination of this number we have
used the five-population-model of interplanetary dust by
Staubach et al. (1997) \cite{staubach97}. So, interstellar particles
should dominate the 
impacts on the CIDA target during cruise, but one has to make sure that an
individual detection is really from an interstellar particle.

We expect the aerogel collector to contain about $120$ interstellar
particles inside its interstellar collection layer. Of these, $40$
should be large particles (size $>1\ {\rm \mu m}$) that are probably
the only ones which are extractable from the aerogel. Due to their
high impact velocities, these large particles might be able to
penetrate very deep into the aerogel and are possibly altered by the
entry process. The distribution of impact angles can be determined by
measuring the geometry of the entry-tracks. This will give valuable
information on the dynamics of interstellar particles in the solar
system. The highest amount of information is of course contained in the
particles themselves when put into the laboratory and analyzed for
their chemical and mineralogical properties. Like CIDA, the collector
might also contain a contamination by interplanetary particles on both
sides, the interstellar and the cometary dust collection layer. As
shown in table \ref{tab_ip} we predict $25$ interplanetary particles
with masses greater than $10^{-13}\;{\rm g}$ to hit the cometary side
of the collector and $20$ to hit the interstellar side.

If the interplanetary particles can be identified as such, it might be
interesting to compare them with the IDPs collected by stratospheric
aircraft \cite{brownlee85,love94}.

\appendix
\section{Spatial Density in the Coma\label{app_a}}

\begin{figure}[htb]
\epsfxsize=.8\hsize\epsfbox{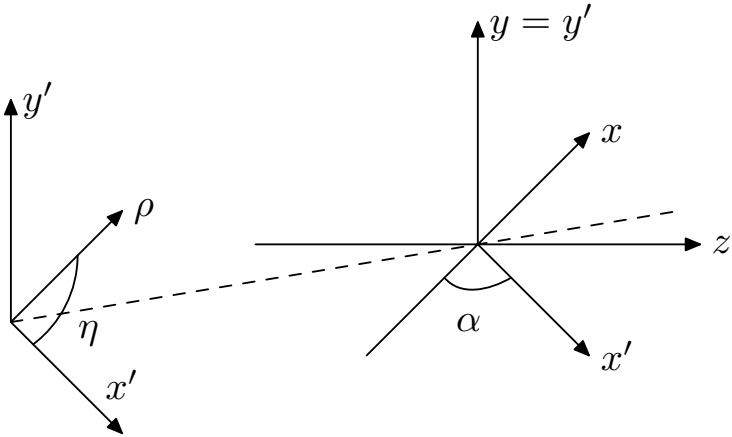}
\caption
{\label{app1} We define the reference frame with the comet in its
origin such that the $z$-axis points in the anti-solar direction and
the observers position is located in the plane spanned by the $x$- and
$z$-axis. In the projection plane, the $y^\prime$-axis is parallel to
the y axis and the $x^\prime$-axis points in the direction of the
projected anti-solar direction. Alternatively to the cartesian
coordinates, a point in the projection plane can be represented by the
cylindrical coordinates $(\rho,\eta)$, where $\rho$ is the distance of
the point from the projected comet position and $\eta$ the angle from
the anti-solar direction.}
\end{figure}

The number of particles which leave the coma per unit time and solid
angle is given by $Z_{e,i}(\vartheta)$, where $\vartheta$ denotes the
angle between ejection and Sun direction. The velocity of the
particles is given by $V_{e,i}(\vartheta)$. For the computation we
introduce a reference frame with the comet nucleus in its origin: The
z-axis is pointing to the anti-solar direction. The observers
position is inside the z-x plane. The position of a particle which was
released in the $(\vartheta,\varphi)$ direction before time $t$ with
respect to this reference frame is given by (see figure \ref{app1}):
\begin{eqnarray}
\vec r (t,\vartheta,\varphi) & = & r
\left(\begin{array}{c}
\cos(\varphi ) \sin(\vartheta )\\
-\sin(\varphi) \sin(\vartheta )\\
-\cos(\vartheta )
\end{array}\right)
\hspace{0.5cm}\\
&& \mbox{with}
\hspace{0.5cm}
r=tV_{e,i}(\vartheta)\nonumber
\end{eqnarray}
The phase angle of the observation is denoted by $\alpha$.
The projected position of a dust particle is given by
(compare figure \ref{app1}):
\begin{equation}\label{proj}
\renewcommand{\arraystretch}{1.5}
\begin{array}{l}
\left(\begin{array}{c} x' \\ y' \end{array} \right)
 = \rho 
\left(\begin{array}{c} \cos(\eta) \\ \sin(\eta) \end{array} \right)
 \\
\hspace{1cm} = r\left(\begin{array}{c}
-\cos(\alpha) \cos(\varphi) \sin(\vartheta) - \sin(\alpha) \cos(\vartheta) \\
-\sin(\varphi) \sin(\vartheta)
\end{array}\right)
\end{array}
\renewcommand{\arraystretch}{1}
\end{equation}
The x-axis in the image plane is the projected anti-solar direction.
Alternatively to the cartesian coordinates in the projected plane 
$(x',y')$ a point in the image plane can be represented by the cylindrical
coordinates $(\rho,\eta)$, where the $\eta=0$ direction points towards 
the (projected) anti-solar direction. The number of particles which are
ejected in the time interval $dt$ and with solid angles inside $d\varphi$
$\sin(\vartheta)d\vartheta$ is given by:
\begin{eqnarray}
dN & =& Z_{e,i}(\vartheta)dt \, d\varphi \, \sin(\vartheta) d\vartheta
\nonumber \\ 
& = &
\frac{Z_{e,i}(\vartheta)}{V_{e,i}(\vartheta)} dr \, d\varphi
\, \sin(\vartheta) d\vartheta
\end{eqnarray}
Next, we determine the contribution of a cone of given angle
$\vartheta$ to the projected number density is computed. The total
number density is then obtained by integrating over all cones in a
second step.

For computing the number of particles inside an infinitesimal area element
in the projected plane $dA=\rho \, d\rho \, d\eta$ due to particles
of the cone $\vartheta$, one needs to know for which radial distance $r$ and
azimuthal angle $\varphi$ there are contributions to the chosen 
point in the image plane $(\rho,\eta)$. This can be found by solving
equation (\ref{proj}) for $r$ and $\cos(\varphi)$ 
(with constant $\vartheta$).
This leads to
\begin{eqnarray}\label{rsolv}
r_{1,2} & = & \frac{\rho}{\cos^2(\vartheta)-\cos^2(\alpha)}
\left(-\cos(\eta)\sin(\alpha)\cos(\vartheta)\right.\nonumber\\
&& \left.\pm \cos(\alpha)
\sqrt{u^2-\cos^2(\vartheta)}\right)\\
\label{phisolv}
\cos(\varphi)_{1,2} & = &\frac{1}{u^2\sin(\vartheta)}
\left(
-\frac{\sin(2\alpha)\cos(\vartheta)}{2}\sin^2(\eta) \right.\nonumber\\
&& \left. \pm
\cos(\eta)\sqrt{u^2-\cos^2(\vartheta)}
\right),
\end{eqnarray}
where $u=u(\alpha,\eta)=\sqrt{1-(\sin(\eta)\sin(\alpha))^2}$.  For
$u<|\cos(\vartheta)|$ the cone $\vartheta$ does not contribute to the
point with angle $\eta$ in the image plane. Furthermore, since
$r=V_{e,i}(\vartheta) t > 0$, only the solutions with $r>0$ have to be
considered. The number of solutions for a given angle $\eta$, phase
angle $\alpha$, and cone $\vartheta$ is denoted by
$N(\alpha,\eta,\vartheta)$ and is given for different conditions in table
\ref{naet}.

Finally, we calculate which infinitesimal azimuthal-radial element
$dr \, d\varphi $ corresponds to the chosen surface element 
$dA=\rho\, d\rho \, d\eta$ in the image plane. This relation is given
by the determinant of the Jacobian matrix:
\begin{equation}
\begin{array}{ccl}
dN&=&
\frac{Z_{e,i}(\vartheta)}{V_{e,i}(\vartheta)} 
\cdot \det \left(
\begin{array}{cc}
\frac{\partial r}{\partial \rho}&
\frac{\partial r}{\partial \eta}\\
\frac{\partial \varphi}{\partial \rho}&
\frac{\partial \varphi}{\partial \eta}
\end{array}
\right) 
\, d\rho \, d\eta
\, \sin(\vartheta) d\vartheta\\
\\
&=&
\frac{Z_{e,i}(\vartheta)}{V_{e,i}(\vartheta)} 
\cdot \left(
\frac{\partial r}{\partial \rho}
\cdot
\frac{\partial \varphi}{\partial \eta}
\right) 
\, d\rho \, d\eta
\, \sin(\vartheta) d\vartheta
\end{array}
\end{equation}
Computation of the partial derivatives yields
\begin{equation}
dN=\frac{Z_{e,i}(\vartheta)}{V_{e,i}(\vartheta)}
\frac{1}{\sqrt{u^2-\cos^2(\vartheta)}}
d\rho \, d\eta \, \sin (\vartheta) d\vartheta.
\end{equation}
Since the last equation gives only the contribution to the number density
due to a single solution of equations (\ref{rsolv}) and (\ref{phisolv}) the
contribution of all solutions has to be summed up to find the total
number of particles in $dA=\rho\, d\rho \, d\eta$ due to cone $\vartheta$:
\begin{equation}
\frac{dN}{dA}=N(\alpha,\eta,\vartheta)
\frac{Z_{e,i}(\vartheta)}{\rho V_{e,i}(\vartheta)}
\frac{1}{\sqrt{u^2-\cos^2(\vartheta)}} \sin (\vartheta) d\vartheta
\end{equation}
Finally, the projected number density is found by integrating over all
cones which contribute to the chosen point $(\rho,\eta)$:
\begin{eqnarray}
\frac{d n_{d,i}}{d(x',y')}(\rho,\eta)& = & \int_{-u}^{u}
N(\alpha,\eta,\vartheta)
\frac{Z_{e,i}(\vartheta)}{\rho V_{e,i}(\vartheta)}\nonumber\\
&&\cdot\frac{1}{\sqrt{u^2-\cos^2(\vartheta)}} d\cos(\vartheta)
\end{eqnarray}
For the calculation of the intensity which is received from the
dust coma the number of particles inside a circular field of
view with radius $\rho$ is needed and can be found by integration of
the projected number density over the field of view (note that
$(N(\alpha,\eta,\vartheta)+N(\alpha,\pi-\eta,\vartheta))/2=1$):
\begin{eqnarray}\label{naper}
N_{d,i}(\rho) & = &
\rho 
\int_{0}^{2\pi}
\int_{-u}^{u}
\frac{Z_{e,i}(\vartheta)}{V_{e,i}(\vartheta)}\nonumber\\
&& \cdot \frac{1}{\sqrt{u^2-\cos^2(\vartheta)}} d\cos(\vartheta)\;
d\eta
\end{eqnarray}

\section{Spatial Density of Hyperbolic Particles\label{app_b}}

\newcommand{\nn}{n_{\infty}}
\newcommand{\vv}{V_{\infty}}

\begin{figure}[htb]
\epsfxsize=.5\hsize\epsfbox{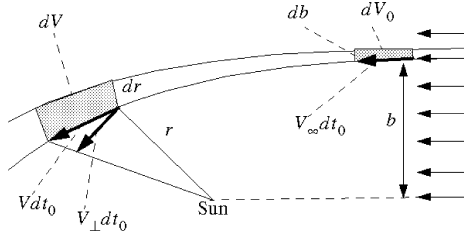}
\caption{\label{app2}
The sketch shows the volume element which contains the particles 
originating from the volume element $dV_0$. The particles on the right
boundary of the volume elements reach the left boundary in the time 
$dt_0$.}
\end{figure}

We derive the number density of particles, which initially enter the
Solar System on parallel
trajectories with velocity $\vv$ and number density $\nn$,  assuming that
the particles are moving on Keplerian trajectories.
For this purpose an infinitesimal volume element $dV$ at distance
$r$ from the Sun and angle $\vartheta$ with respect to the direction
the particles enter the Solar System is considered (see figure
\ref{app2}).
The particles which cover the volume element $dV$ has started in the
time interval $dt_0$ from outside the impact parameter interval $db$
and covered 
at a large distance from the Sun the volume element 
$dV_0=2\pi b \, db \, \vv dt_0$. 
The number of particles inside both volume
elements is given by $dN=\nn \, dV_0$.
The size of the volume element $dV$ can be computed by
\begin{equation}
dV=2\pi r \sin(\vartheta) \, dr \, V_{\bot} \, dt_0 =
2\pi r \, \frac{dr}{db} db \, V_{\bot} \, dt_0.
\end{equation}
Due to the preservation of the angular momentum on a Keplerian 
trajectory the velocity perpendicular to the radial direction
can be expressed by $V_{\bot}=b\vv/r$ and the size of the volume
element can be written
\begin{equation}
dV=\sin(\vartheta)\frac{dr}{db} \,d V_0.
\end{equation}
Thus, only the derivative of the radial distance $r$ with respect to
the impact parameter needs to be computed. The radial distance $r$ of
a non radial Keplerian trajectory can be parameterized:
\begin{equation}\label{rstd}
r=\frac{p}{1+e \cos (\vartheta-\vartheta_0)},
\end{equation}
where $p$, $e$ and $\vartheta_0$ are constants which has to be expressed
by the initial condition:
\begin{eqnarray}
p & = & \frac{b^2}{d} \nonumber\\
e & = & \pm \sqrt{1+\left( \frac{b}{d} \right)^2}\nonumber\\
\cos(\vartheta_0)&=&-\frac{1}{e}\nonumber\\
\sin(\vartheta_0) & = & \frac{b}{de}\nonumber\\
d & = & \frac{\mu}{\vv^2},
\end{eqnarray}
where $\mu=\gamma M_{\odot}(1-\beta)$ is the effective gravitational
constant of the Sun and $e$ is positive in the case $\mu>0$ and
negative for $\mu<0$. By inserting the constants $p$, $e$ and
$\vartheta_0$ in equation (\ref{rstd}), we find the radial distance as
a function of the impact parameter $b$:
\begin{equation}\label{rnstd}
r(b)=\frac{b^2}{d}\cdot
\frac{1}{1-\cos(\vartheta)+\frac{b}{d}\sin(\vartheta)}
\end{equation}
By computing the derivative with $b$ an expression for the number 
density at the position $(r,\vartheta)$ is found:
\begin{equation}\label{nsngsol}
n=\frac{dN}{dV}=
\nn\cdot\frac{b^2}{r \sin(\vartheta) |2b-r\sin(\vartheta)|}
\hspace{1cm}
\vartheta>0
\end{equation}
For computing the number density at a given position in the Solar System,
the impact parameter $b$ as a function of $r$ and $\vartheta$ is needed,
what can be found by solving equation (\ref{rnstd}) for $b$:
\begin{equation}
b=\left|\frac{r\sin(\vartheta)}{2}\pm
\sqrt{\left(\frac{r\sin(\vartheta)}{2}\right)^2+rd(1-\cos(\vartheta))}
\right|
\end{equation}
For an attractive effective gravitational constant $\mu>0$, i.e. $d>0$,
every position can be reached by two trajectories
(apart from the axes $\vartheta=0$ and $\vartheta=\pi$). 
For a repulsive effective gravitational
constant $\mu<0$ i.e. $d<0$ every position outside the parabola 
$|r \sin(\vartheta)|=4\sqrt{d^2+d/2 \cdot r\cos(\vartheta)}$ is reached
by two different trajectories, no trajectories reach the interior
of the parabola and every place on the parabola is reached by exactly 
one trajectory. Since equation (\ref{nsngsol}) gives only the 
contribution to the number density of one solution the
contributions due to the different solutions
which reach a point has to be added up to calculate the total number
density at a point. On the axis $\vartheta=\pi$ for $\mu>0$ and
on the parabola for $\mu<0$ the volume element $dV$ becomes zero.
Therefore the number density becomes infinite at those positions. However,
the infinities are of that kind that any integral of the number
density over a finite volume is finite. 

Because the representation of a Keplerian orbit as it is given in
equation (\ref{rstd}) cannot be applied to radial trajectories
i.e. impact parameter $b=0$ and $\vartheta=0$, the number density
at the axis $\vartheta=0$ cannot be computed directly by equation 
(\ref{nsngsol}), but the formula must be smoothly continued to this axis:
\begin{equation}
n=\nn \frac{1}{4}\left(\frac{1}{\sqrt{1+\frac{2d}{r}}}+
\sqrt{1+\frac{2d}{r}}\pm 2\right),
\end{equation}
The upper sign applies to the contribution to the number
density due to the particles approaching the Sun
and the lower sign to the departing particles.
\end{document}